\documentclass[aps,twocolumn,pra, showpacs,superscriptaddress]{revtex4-1}
\usepackage{graphicx}
\usepackage{amsmath}
\usepackage{amssymb}
\usepackage{pifont}
\usepackage[colorlinks,citecolor=blue]{hyperref}

\begin{document}

\title{Circuit analog of quadratic optomechanics} 

\author{Eun-jong Kim}
\email{vb777@snu.ac.kr}
\affiliation{Department of Physics and Astronomy, Seoul National University, Seoul, 151-747 Korea}
\affiliation{iTHES Research Group, RIKEN, Wako-shi, Saitama, 351-0198 Japan}
\author{J.R. Johansson}
\email{robert@riken.jp}
\affiliation{iTHES Research Group, RIKEN, Wako-shi, Saitama, 351-0198 Japan}
\author{Franco Nori}
\affiliation{CEMS, RIKEN, Wako-shi, Saitama, 351-0198 Japan}
\affiliation{Department of Physics, University of Michigan, Ann Arbor, Michigan
48109-1040 USA}

\begin{abstract}
We propose a superconducting electrical circuit that simulates a quadratic optomechanical system. A capacitor placed between two transmission-line (TL) resonators acts like a semi-transparent membrane, and a superconducting quantum interference device (SQUID) that terminates a TL resonator behaves like a movable mirror. Combining these circuit elements, it is possible to simulate a quadratic optomechanical coupling whose coupling strength is determined by the coupling capacitance and the tunable bias flux through the SQUIDs. Estimates using realistic parameters suggest that an improvement in the coupling strength could be realized, to five orders of magnitude from what has been observed in membrane-in-the-middle cavity optomechanical systems. This leads to the possibility of achieving the strong-coupling regime of quadratic optomechanics.
\end{abstract}

\date{\today}
\pacs{42.50.Pq, 42.50.Wk, 85.25.Cp}
\maketitle

\section{Introduction}\label{sec:intro}
Optomechanics is the study of interactions between optical and mechanical degrees of freedom \cite{Aspelmeyer:2012fy, Meystre:2013gu, aspelmeyer:2013}. It has been a burgeoning field in recent years, with various theoretical proposals and experimental realizations (e.g., sideband cooling of mechanical oscillators to quantum ground states \cite{OConnell:2010br, Teufel:2011jg, Chan:2011dy} and normal-mode splitting \cite{Groblacher:2009eh,Teufel:2011ih, Verhagen:2012ei}). In these works, the interaction Hamiltonian is linear in the displacement of the mechanical oscillator. Another type of interaction, which is quadratic in the displacement of the mechanical oscillator, has also been demonstrated \cite{Thompson:2008dx, Jayich:2008iz, Sankey:2009vs, Sankey:2010ej}, stimulating other theoretical proposals \cite{Bhattacharya:2008kd, Clerk:2010ht, Nunnenkamp:2010gj, Liao:2013fs, Liao:2014ev}. In particular, quadratic optomechanics opened up the possibility of quantum nondemolition (QND) measurements \cite{Braginsky:1980qv} of the mechanical oscillator's energy eigenstates. However, it has been suggested \cite{Miao:2009jk} that a strong quadratic coupling strength is required to resolve a single mechanical quantum.

Meanwhile, the field of circuit quantum electrodynamics (cQED) emerged as a promising candidate for future quantum information processing \cite{You:2005gn, Wendin:2006, Schoelkopf:2008cs, Clarke:2008gi, Devoret:2013jz, Xiang:2013hm}. Josephson junction-based devices together with transmission-line (TL) resonators have proved effective in the manipulation and the readout of superconducting qubits.
Also, cQED has drawn attention as an analog system for probing various quantum phenomena \cite{Buluta:2009ii, You:2011uv, Nation:2012ka, Georgescu:2014bg}. In particular, superconducting quantum interference devices (SQUIDs) can implement tunable boundary conditions in circuits. With this principle, SQUIDs have been employed as a method for introducing \emph{in-situ} tunability to circuits \cite{Wallquist:2006fc, Sandberg:2008br, PalaciosLaloy:2008ef, CastellanosBeltran:2008cg, Wustmann:2013bm}, demonstrating physical effects that had not previously been observed, e.g., the dynamical Casimir effect (DCE) \cite{Johansson:2009zz, Johansson:2010bx, Wilson:2011wr, Johansson:2013tk}. Other theoretical proposals for analog circuit realizations include Hawking radiation \cite{Nation:2009jc}, entanglement of superconducting qubits using DCE \cite{Felicetti:2014dz}, and the twin paradox \cite{Lindkvist:2014hc}. Also, an all-circuit realization of standard linear optomechanics has recently been proposed \cite{johansson:2014}.

\begin{figure}[b]
	\includegraphics[width=0.48\textwidth]{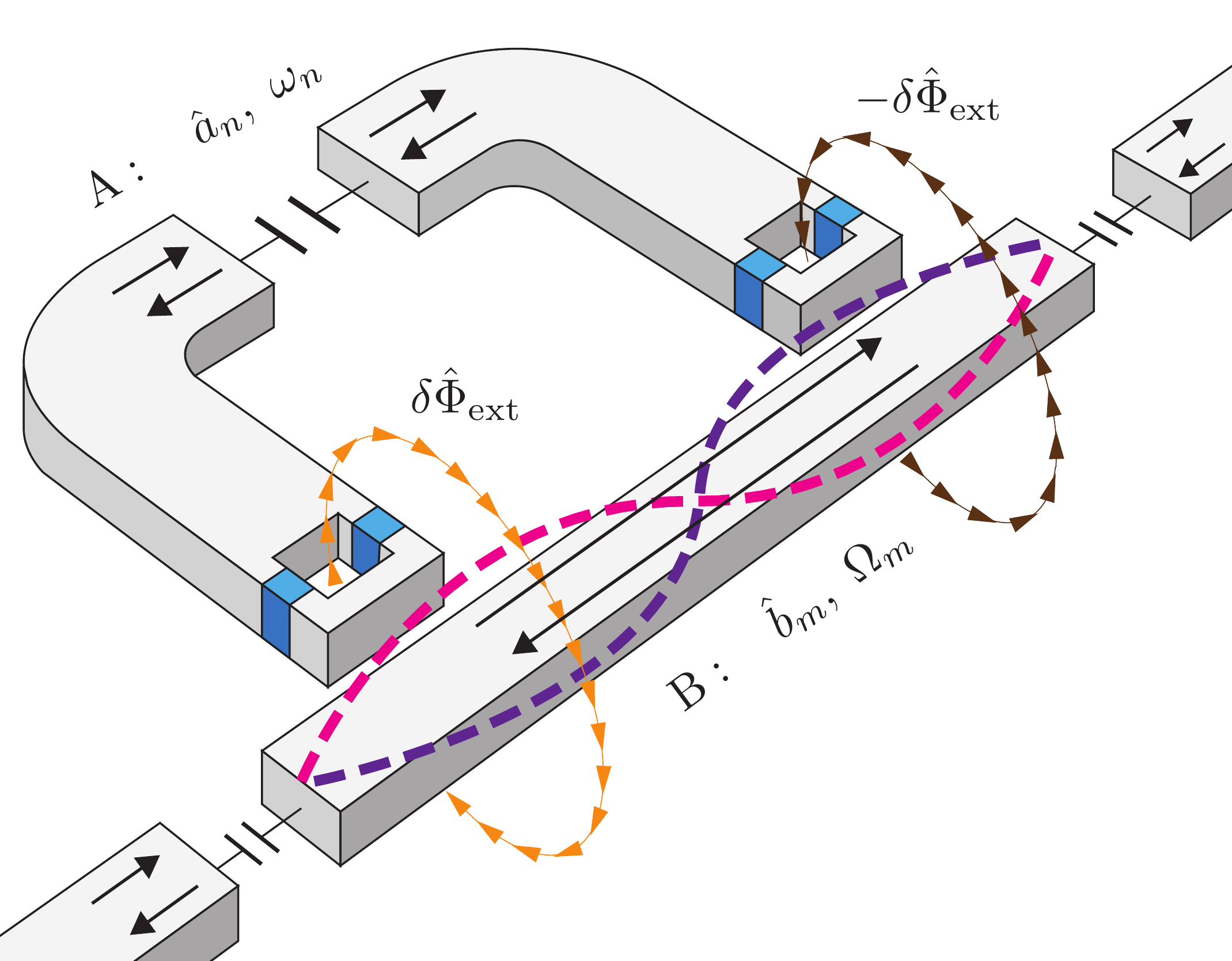}
	\caption{(color online) Schematic design of an analog circuit for quadratic optomechanics. Resonator A, described by the annihilation operator $\hat{a}_n$ and the mode frequency $\omega_n$, consists of two capacitively-coupled SQUID-terminated TL resonators. Resonator B, described by the annihilation operator $\hat{b}_m$ and the mode frequency $\Omega_m$, is a TL resonator. Resonator A and resonator B provide optical and pseudo-mechanical degrees of freedom, respectively. The current distribution of resonator B is chosen to be antisymmetric to ensure opposite flux variations $\pm \delta\hat{\Phi}_\mathrm{ext}$ through the SQUIDs forming resonator A.}
	\label{fig:intro-schematic}
\end{figure}

In this paper, we present a superconducting electrical circuit, illustrated in Fig.~\ref{fig:intro-schematic}, that simulates a quadratic optomechanical system. The system consists of two resonators, denoted as resonator A and resonator B, each corresponding to the optical cavity and the mechanical oscillator of quadratic optomechanics. The coupling capacitor and the SQUIDs forming resonator A correspond to a fixed semi-transparent membrane and movable optical cavity ends, respectively. By synchronizing the motion of the movable cavity ends, which is accomplished by applying opposite flux variations through the SQUIDs of resonator A, a relative displacement of the fixed membrane with respect to the cavity center is generated. Due to this parametrically-induced frequency shift of resonator A, the position quadrature of resonator B couples quadratically to the photon number of resonator A in a certain regime. Although the physics underlying cavity quadratic optomechanics and our circuit proposal is \mbox{intrinsically} different, the interaction is of the same form.

The remaining part of this paper is outlined as follows:
Section~\ref{sec:review} reviews the basic principles of quadratic optomechanical systems \cite{Thompson:2008dx, Jayich:2008iz, Sankey:2009vs, Sankey:2010ej} that are employed in our discussion.
In Sec.~\ref{sec:circuitmodel}, we investigate the mode frequencies and the mode structures of specific circuit models to find a circuit analog of optical and mechanical elements.
In Sec.~\ref{sec:hamiltonian}, the quantization procedure of the system as well as Hamiltonian formulation of our analog quadratic optomechanical system is presented. In Sec.~\ref{sec:circuit-realization}, we suggest that the proposed circuit is realizable, potentially giving rise to a large improvement in the quadratic coupling strength compared to cavity-optomechanical systems. The summary of our results follows in Sec.~\ref{sec:conclusion}.

\section{Review of quadratic optomechanics}\label{sec:review}
Quadratic optomechanical coupling was first demonstrated in Ref.~\cite{Thompson:2008dx}. This system consists of an optical cavity partitioned by a semi-transparent membrane. The basic idea of this system is that the mode frequency of the cavity $\omega_\mathrm{cav}$ as a function of membrane displacement $\xi$ from the cavity center has local extrema, where the first-order derivatives $\omega'_\mathrm{cav}(\xi)$ vanish. To be specific, defining $v$ as the speed of light inside the cavity, the mode frequencies $\omega_\mathrm{cav}(\xi) = kv$ satisfy \cite{Jayich:2008iz}
\begin{align}
\cos{(k d - \delta)} = |r|\cos{(2k\xi)},\label{eq:eigenmode-eq}
\end{align}
where $d$ is the total length of the cavity, $\delta$ is the overall phase, and $r$ is the reflectivity of the membrane which is close to unity. Choosing the extremum point $\xi=0$ as the center of oscillation, the Hamiltonian is written in the form
\begin{align}
\hat{H} = \hbar\omega_\mathrm{cav}(0) \hat{a}^\dagger \hat{a} +  \hbar \Omega \hat{b}^\dagger \hat{b} -\hbar g \hat{a}^\dagger \hat{a} (\hat{b}^\dagger + \hat{b})^2,
\end{align}
where $\Omega$ is the mechanical oscillation frequency of the membrane and $g = {\hbar}\omega''_\mathrm{cav}(0)/{4m \Omega}$ is the quadratic coupling strength ($m$ is the mass of the membrane). Here, $\hat{a}$ and $\hat{b}$ denote the annihilation operators for the optical mode of the cavity (photon) and the mechanical mode of the membrane (phonon), respectively.

This system distinguishes itself from the standard linear optomechanical system \cite{Aspelmeyer:2012fy, Meystre:2013gu, aspelmeyer:2013} in several respects: $(i)$ since the cavity ends remain fixed and the membrane possesses a mechanical degree of freedom, experimentalists can circumvent the difficulty of combining high-finesse cavities with mechanical degrees of freedom; $(ii)$ neglecting fast-oscillating terms, the quadratic coupling part of the Hamiltonian $\hat{a}^\dagger \hat{a}(\hat{b}^\dagger+\hat{b})^2$ reduces to $2\hat{a}^\dagger \hat{a}\hat{b}^\dagger \hat{b}$, which enables QND phonon number measurements \cite{Braginsky:1980qv} of the mechanical oscillator, since $[\hat{H}, \hat{b}^\dagger \hat{b}]=0\:$; $(iii)$ by choosing the membrane displacement $\xi$ such that the first-order derivative does not vanish, the system returns to the linear optomechanics regime. 

In general, the position-squared sensitivity $\omega_\mathrm{cav}''$ of the cavity frequency of Ref.~\cite{Thompson:2008dx} is too small to achieve the QND phonon number readout \cite{Sankey:2009vs}. Using the parameters $L=6.7\ \mathrm{cm}$, $r = 0.999$, $\lambda = 532\ \mathrm{nm}$, $m=50\ \mathrm{pg}$, and $\Omega/2\pi = 100\ \mathrm{kHz}$, in Ref.~\cite{Thompson:2008dx},
\begin{align*}
\omega_\mathrm{cav}'' =  \frac{16\pi^2 c}{L \lambda^2} \sqrt{2(1-r)} \approx 2\pi \times 18\ \mathrm{kHz \ nm^{-2}},
\end{align*}
and the ratio of the quadratic coupling strength $g$ to the mechanical mode frequency $\Omega$ is given by
\begin{align}
\frac{g}{\Omega} = \frac{\hbar\omega_\mathrm{cav}''}{4m\Omega^2}=9.4\times 10^{-13}.\label{eq:coupling-strength-cavity-1}
\end{align}

In Ref.~\cite{Sankey:2010ej}, an angular degree of freedom, i.e., tilt of the membrane, was introduced as a method of increasing the quadratic coupling strength. If the system is perfectly symmetric, transverse modes of the cavity (for example, $\mathrm{TEM_{\{20, 11, 02\}}}$) are degenerate. On the other hand, when this system has an asymmetry, either due to a tilt of the membrane or an imperfection of the cavity, the mode degeneracy is lifted to give additional local extrema of the cavity frequency with larger values of the second-order derivatives $\omega_\mathrm{cav}''$. This may increase $\omega_\mathrm{cav}''$ by three orders of magnitude.

From the parameters in Ref.~\cite{Sankey:2010ej}, $\Omega /2\pi = 100\ \mathrm{kHz}$, $m=50\ \mathrm{pg}$, and $\omega_\mathrm{cav}''/2\pi = 10\ \mathrm{MHz\ nm^{-2}}$, the ratio of the coupling strength $g$ to the mechanical oscillation frequency $\Omega$ is estimated as:
\begin{align}
\frac{g}{\Omega} = \frac{\hbar\omega_\mathrm{cav}''}{4m\Omega^2}=5.3\times 10^{-10}.\label{eq:coupling-strength-cavity-2}
\end{align}
Still, the coupling strength is very small compared to the mode frequencies of the cavity and the mechanical oscillator.

In general, it has been an experimental challenge in cavity-optomechanical systems to reach a quadratic coupling strength high enough to achieve QND measurements of the phonon number \cite{Miao:2009jk, Clerk:2010ht}. As an alternative approach for exploring quadratic optomechanics, Bose-Einstein condensate (BEC) systems have previously been proposed and demonstrated \cite{Brennecke:2008el, Murch:2008kj, Purdy:2010gh, Jing:2011ie}. Here, we look for an analog in cQED to possibly realize strong quadratic coupling strengths.

\begin{figure*}[t]
	\includegraphics[width=\textwidth]{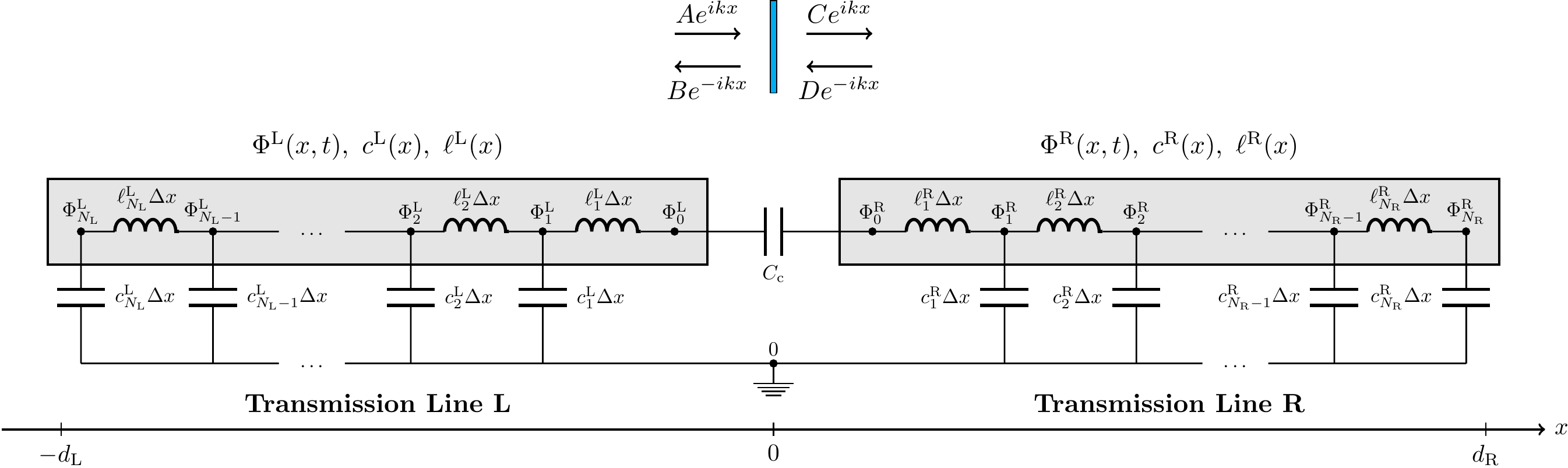}
	\caption{(color online) Two capacitively-coupled TL resonators expressed as a lumped-element circuit. The TL resonators on the left and the right are labeled with $\alpha=\mathrm{L}$ and $\alpha=\mathrm{R}$, respectively. Each TL resonator can be modeled as an infinite number of LC circuits, each with node capacitance $c_k^{\alpha}\Delta x$ and node inductance $\ell_k^{\alpha} \Delta x$ ($1\le k\le N_{\alpha}$). In the continuum limit, the discrete node flux $\Phi^{\alpha}_k(t)$, node capacitance per unit length $c_k^{\alpha}$, and node inductance per unit length $\ell_k^{\alpha}$, converge to continuous functions inside each TL resonator $\Phi^{\alpha}(x,t)$, $c^{\alpha}(x)$, and $\ell^{\alpha}(x)$, respectively. In the middle, there is a capacitor $C_\mathrm{c}$ which couples the two TL resonators. If both TL resonators are uniform and homogeneous, this capacitor can be thought of as a partially-transparent membrane (shown in blue) giving rise to a linear transformation between the wave amplitudes of different regions.}
	\label{fig:capcoupling-schematic}
\end{figure*}

\section{Circuit model}\label{sec:circuitmodel}
In this section, we discuss how optical and mechanical elements can be mapped onto circuit elements. The eigenmode equation Eq.~\eqref{eq:eigenmode-eq} that we observe in the standard fixed ``membrane-in-the-middle'' optical system is the same as our ``capacitor-in-the-middle'' TL resonator configuration in Sec.~\ref{sec:capcoupling}. In Sec.~\ref{sec:squidres}, we look at how SQUID-terminated TL resonators can introduce a variable length of the resonator, which offers tunability of the resonance frequency. Section~\ref{sec:analogcircuit} combines the two principles to simulate a movable membrane in the middle of the resonator whose position can be adjusted by an external flux.

\subsection{Capacitively-coupled resonators}\label{sec:capcoupling}
We first discuss capacitively-coupled TL resonators, as depicted in Fig.~\ref{fig:capcoupling-schematic}.
We define $\Phi^\alpha(x,t)\equiv \int_{-\infty}^{t}V^{\alpha}(x,t') \: \mathrm{d}t'$ as the flux field, and $c^{\alpha}(x)$ and $\ell^{\alpha} (x)$ are the characteristic capacitance and inductance per unit length at position $x$ and time $t$ of a TL resonator ($\alpha=\mathrm{L}, \mathrm{R}$). Then, the Lagrangian of the system can be expressed in terms of the Lagrangian density \cite{yurke:1984, devoret:1995}, $L=\int_{-d_\mathrm{L}}^{d_\mathrm{R}}\mathcal{L}\:\mathrm{d} x$, with
\begin{align*}
\mathcal{L} &= \left\{ \frac{c^\mathrm{L}(x)}{2}\left[\partial_t \Phi^\mathrm{L}(x,t)\right]^2 - \frac{1}{2\ell^\mathrm{L}(x)}\left[\partial_x \Phi^\mathrm{L}(x,t)\right]^2 \right\}\Theta(-x)\\
&\quad+\left\{ \frac{c^\mathrm{R}(x)}{2}\left[\partial_t \Phi^\mathrm{R}(x,t)\right]^2 - \frac{1}{2\ell^\mathrm{R}(x)}\left[\partial_x \Phi^\mathrm{R}(x,t)\right]^2 \right\}\Theta(x)\\
&\quad+\frac{C_\mathrm{c}}{2} \left[\partial_t \Phi^\mathrm{R}(x,t) - \partial_t \Phi^\mathrm{L}(x,t)\right]^2 \delta(x).
\end{align*}
Here, $\delta(x)$ is the one-dimensional Dirac delta function and $\Theta(x)$ is the Heaviside step function. Also, $C_\mathrm{c}$ is the capacitance of the capacitor between the two TL resonators.
Applying the Euler-Lagrange equation of motion \cite{goldstein},
\begin{align}
\frac{\partial\mathcal{L}}{\partial \Phi^\alpha}- \frac{\partial}{\partial x}\frac{\partial\mathcal{L}}{\partial_x \Phi^\alpha}-\frac{\partial}{\partial t}\frac{\partial\mathcal{L}}{\partial_t \Phi^\alpha}=0\quad(\alpha=\mathrm{L},\ \mathrm{R}),\label{eq:euler-lagrange}
\end{align}
we obtain the partial differential equation for $x\neq0$,
\begin{align}
\frac{\partial}{\partial x} \left[ \frac{1}{\ell(x)} \partial_x \Phi(x,t)\right] - c(x) \partial_{tt}\Phi(x,t) = 0, \label{eq:partial-diff-tl}
\end{align}
subject to the boundary conditions at $x=0$,
\begin{align}
C_\mathrm{c} \left[\partial_{tt}\Phi(0^+,t)-\partial_{tt}\Phi(0^-,t)\right] &= \frac{1}{\ell(0^+)}\partial_x \Phi(0^+, t),\label{eq:capcoupling-bd-1} \\
C_\mathrm{c} \left[\partial_{tt}\Phi(0^-,t)-\partial_{tt}\Phi(0^+,t)\right] &= -\frac{1}{\ell(0^-)}\partial_x \Phi(0^-, t).\label{eq:capcoupling-bd-2}
\end{align}
Without loss of generality, we let $f(x) \equiv f^{\mathrm{L}}(x)\Theta(-x) + f^{\mathrm{R}}(x)\Theta(x)$ ($f=\Phi,c,\ell$). Note that adding Eq.~\eqref{eq:capcoupling-bd-1} and Eq.~\eqref{eq:capcoupling-bd-2} yields the current-conservation relation at the boundary,
\begin{align*}
\frac{1}{\ell(0^+)} \partial_x\Phi(0^+,t) = \frac{1}{\ell(0^-)} \partial_x\Phi(0^-,t).
\end{align*}
We consider the special case where both ends of the TL resonators are grounded. We further assume that the TL resonators are homogeneous, having identical characteristic capacitance and inductance per unit length $c(x)=c_0$, $\ell(x)=\ell_0$. In this case, our problem reduces to solving the partial differential equation for $x\neq 0$, ($v_0 \equiv 1/\sqrt{\ell_0 c_0}$)
\begin{align}
\partial_{xx} \Phi(x,t) - \frac{1}{v_0^2} \partial_{tt}\Phi(x,t) = 0, \label{eq:wave-eqn}
\end{align}
which is the massless Klein-Gordon wave equation \cite{goldstein}, subject to the four boundary conditions
\begin{subequations}
\begin{align}
C_\mathrm{c} \left[\partial_{tt}\Phi(0^+,t)-\partial_{tt}\Phi(0^-,t)\right] &= \frac{1}{\ell_0}\partial_x \Phi(0^+, t),\label{eq:capcoupling-bd-3}\\
C_\mathrm{c} \left[\partial_{tt}\Phi(0^-,t)-\partial_{tt}\Phi(0^+,t)\right] &= -\frac{1}{\ell_0}\partial_x \Phi(0^-, t),\label{eq:capcoupling-bd-4}\\
\Phi(-d_\mathrm{L},t)&=0,\label{eq:capcoupling-bd-5}\\
\Phi(d_\mathrm{R},t)&=0.\label{eq:capcoupling-bd-6}
\end{align}
\end{subequations}
We look for a solution of the form $\Phi(x,t) = u(x)\psi(t)$, 
using separation of variables. The wave equation then yields two independent ordinary differential equations,
\begin{align}
\begin{split} 
u''(x) + k^2 u(x) &= 0,\\
\ddot{\psi}(t) + \omega^2 \psi(t) &= 0,
\end{split}\label{eq:two-ode}
\end{align}
where $k$ is a constant, and $\omega=kv_0$. The boundary conditions in our case depend only on $x$, and we only need to solve the ordinary differential equation for $u(x)$. The general solution for $u(x)$ is a linear combination of $e^{\pm ikx}$, with different amplitudes,
$$
u(x) = \begin{cases} Ae^{ikx} + Be^{-ikx} & (x<0), \\
				Ce^{ikx} + De^{-ikx} & (x>0),
	   \end{cases} 
$$
as illustrated in Fig.~\ref{fig:capcoupling-schematic}. The boundary conditions Eqs.~\eqref{eq:capcoupling-bd-3}-\eqref{eq:capcoupling-bd-4}, which correspond to a capacitive coupling, yield a linear, ``fixed membrane''-like transformation between the wave amplitudes:
\begin{align}
\begin{pmatrix} B \\ C \end{pmatrix}
=
\begin{pmatrix} r & it \\ it & r \end{pmatrix}
\begin{pmatrix} A \\ D \end{pmatrix},
 \label{eq:capcoupling-membrane}
\end{align}
where $r$ and $t$ are the effective reflectivity and transmissivity arising from the capacitive coupling. Here,
\begin{align}
r = \frac{i\frac{\omega_\mathrm{c}}{2\omega}}{1+i\frac{\omega_\mathrm{c}}{2\omega}},\quad t = \frac{-i}{1+i\frac{\omega_\mathrm{c}}{2\omega}},
\end{align}
and $\omega_\mathrm{c} \equiv (\sqrt{\ell_0/c_0} C_\mathrm{c})^{-1}$ is the characteristic frequency of the capacitive coupling. Note that the reflectivity and transmissivity satisfy $|r|^2+|t|^2=1$. This transformation, Eq.~\eqref{eq:capcoupling-membrane}, is equivalent to the transformation matrix between the field operators mentioned in Ref.~\cite{Johansson:2010bx}.

Increasing the capacitance $C_\mathrm{c}$ amounts to increasing transmissivity and reducing reflectivity; decreasing $C_\mathrm{c}$, on the other hand, enhances reflectivity while suppressing transmissivity. In the limit $C_\mathrm{c} \rightarrow 0$, the reflectivity approaches unity, corresponding to open-ended (i.e., completely decoupled) boundary condition at $x=0$.

The boundary conditions Eqs.~\eqref{eq:capcoupling-bd-5}-\eqref{eq:capcoupling-bd-6}, which correspond to grounded ends, result in the total reflection of waves at the ends of the TL resonators. This produces a ``mirror''-like transformation between wave amplitudes:
\begin{align}
Ae^{-ikd_\mathrm{L}} + Be^{ikd_\mathrm{L}} = Ce^{ikd_\mathrm{R}} + De^{-ikd_\mathrm{R}} =0. \label{eq:capcoupling-mirror}
\end{align}
Eq.~\eqref{eq:capcoupling-membrane} in tandem with Eq.~\eqref{eq:capcoupling-mirror} impose a constraint on the allowed frequencies of the system, on the form of an optical cavity with a fixed membrane in the middle:
\begin{subequations}
\begin{align}
 \frac{\omega_\mathrm{c}}{\omega_n}= \tan{\left(\frac{\omega_n d_\mathrm{L}}{v_0}\right)} + \tan{\left(\frac{\omega_n d_\mathrm{R}}{v_0}\right)} , \label{eq:capcoupling-mode-1}
\end{align}
or, equivalently,
\begin{align}
\cos{(k_n d-\delta_n)} = |r_n| \cos{(2k_n\xi)}, \label{eq:capcoupling-mode-2}
\end{align}
\end{subequations}
where $k_n = \omega_n/v_0$. Here, $n$ is used to label the discrete modes and we have introduced the total length of the cavity $d = d_\mathrm{L} + d_\mathrm{R}$, the displacement $\xi = (d_\mathrm{L}-d_\mathrm{R})/2$ of the capacitor from the center, and the phase angle $\delta_n$, which satisfies
\begin{align*}
\cos{\delta_n} =- |r_n|,\qquad
\sin{\delta_n} = |t_n|.
\end{align*}
Note that Eq.~\eqref{eq:capcoupling-mode-2} is identical to the eigenmode equation Eq.~\eqref{eq:eigenmode-eq} of cavity quadratic optomechanics.
Equation~\eqref{eq:capcoupling-mode-2} makes it possible to expand the normal-mode frequencies in the displacement parameter $\xi$,
\begin{align}
\omega_n(\xi)=\omega_n^{(0)}+\omega_n^{(2)}\xi^2 +\omega_n^{(4)}\xi^4 +\mathcal{O}(\xi^6),\label{eq:normalmodefreq-asymptote}
\end{align}
where the expansion coefficients are given by $(n=0, 1, 2, \ldots)$
\begin{align*}
\omega_n^{(0)} &= \frac{\pi v_0}{d}\bigg( n+\textrm{mod}(n+1,2)\bigg) &\\
			&\quad -\frac{2v_0\cos{^{-1}(|r_n^{(0)}|)}}{d}\textrm{mod}(n+1,2),\\
\omega_n^{(2)} &= -\frac{(-1)^n}{d}\frac{\omega_n^{(0)}\omega_\mathrm{c}}{v_0},\\
\omega_n^{(4)} &=
\frac{(-1)^n}{d}\frac{\omega_n^{(0)} \omega_\mathrm{c}^3}{12v_0^3}\left(1+\frac{4({\omega_n^{(0)}})^2}{\omega_\mathrm{c}^2}\right).
\end{align*}
Here, $\textrm{mod}(n+1,2)$ is the modulus function that returns $1$ for even values of $n$, and $0$ for odd values of $n$. Note that the expansion coefficient for the first and the third order is zero, i.e., $\omega^{(1)}_n=\omega^{(3)}_n=0$. Therefore, we observe a quadratic dependence of normal-mode frequencies $\omega_n$ on the displacement parameter $\xi$, up to third order.

In the rest of our discussion, we use the third-order expansion, $\omega_n(\xi)\approx \omega_n^{(0)} + \omega_n^{(2)}\xi^2$, as an approximate analytic expression. To quantify the validity of this approximation, we introduce a new parameter called a \emph{validity extent}. The third-order approximation is accurate to 99\% in the range $|\xi|\le {\xi}_{n*}$, where  ${\xi}_{n*}$ is given by
\begin{align}
{\xi}_{n*} \equiv \frac{v_0}{\sqrt[4]{{\omega_n^{(0)}\omega_\mathrm{c}^3} + 4{\left(\omega_n^{(0)}\right)^3 \omega_\mathrm{c}}}} \cdot (12 \times 10^{-2})^{1/4}.\label{eq:validity-range}
\end{align}
Thus, the region of $\xi$ where this third-order approximation holds is larger for lower modes and stronger capacitive coupling, and smaller for higher modes and weaker capacitive coupling.

\begin{figure}
  \includegraphics[width=.48\textwidth]{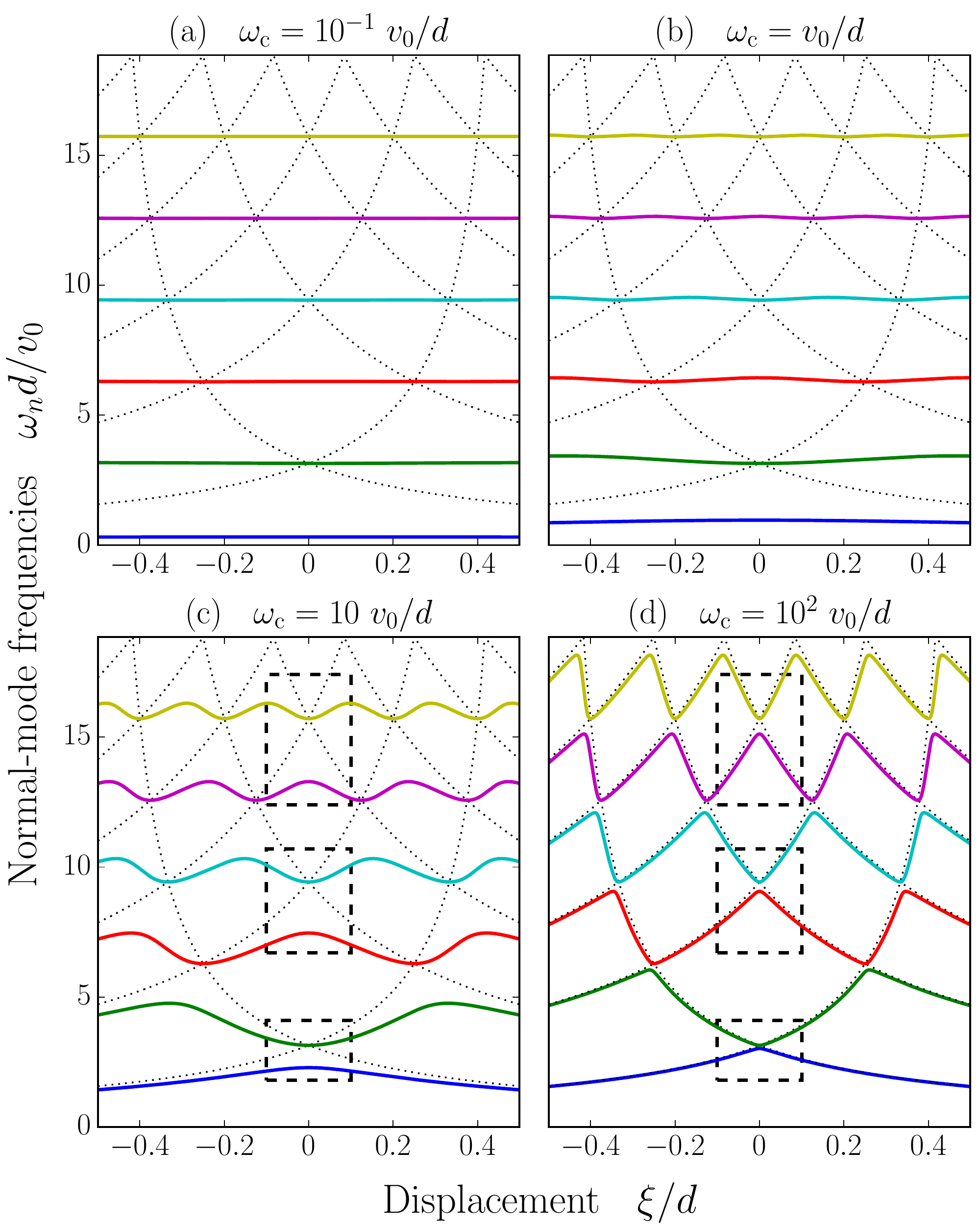}
  \caption{(color online) Normal-mode frequencies of the first six modes ($n=0$ to $n=5$, bottom to top), calculated from Eq.~\eqref{eq:capcoupling-mode-1}. The four panels show the mode frequencies, as a function of the location of the capacitor, for decreasing coupling strengths: (a) $\omega_\mathrm{c}=10^{-1}\ v_0/d$, (b) $\omega_\mathrm{c}= v_0/d$, (c) $\omega_\mathrm{c}=10\ v_0/d$, (d) $\omega_\mathrm{c}=10^{2}\ v_0/d$. The dotted lines are the normal-mode frequencies for the completely-decoupled case ($C_\mathrm{c}=0$ or $\omega_\mathrm{c}\rightarrow\infty$). The regions inside the dashed boxes are zoomed in Fig.~\ref{fig:capcoupling-normalmodefreq-enlarged}.}
	\label{fig:capcoupling-normalmodefreq}
\end{figure}
\begin{figure}
  \includegraphics[width=.48\textwidth]{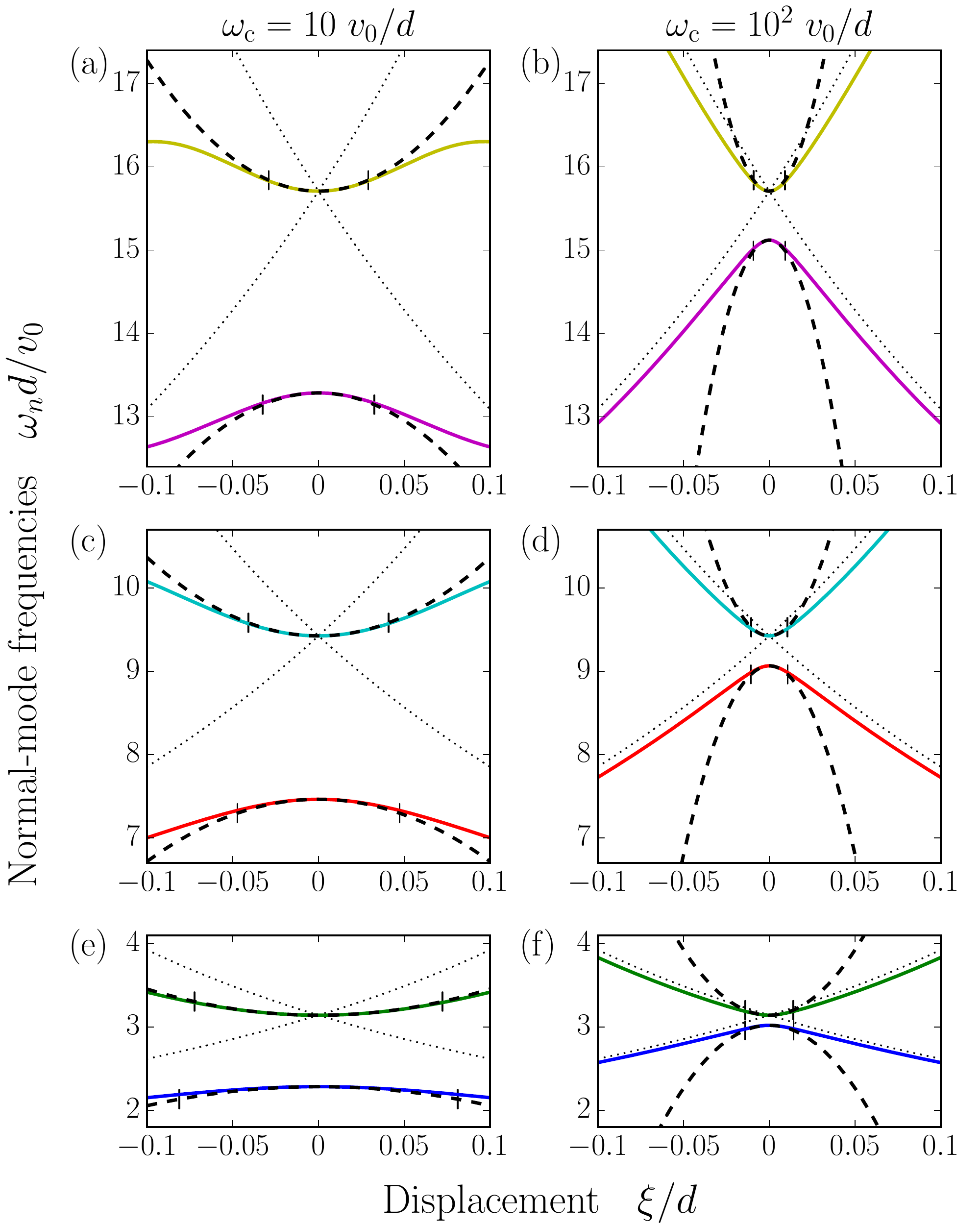}
  \caption{(color online) Enlargement (with the same scale) of the dashed boxes in Fig.~\ref{fig:capcoupling-normalmodefreq}(c) and Fig.~\ref{fig:capcoupling-normalmodefreq}(d). Here, (a), (c), (e) and (b), (d), (f) correspond to the dashed boxes in Fig.~\ref{fig:capcoupling-normalmodefreq}(c) and Fig.~\ref{fig:capcoupling-normalmodefreq}(d), from top to bottom, respectively.  The dashed lines are plotted with Eq.~\eqref{eq:normalmodefreq-asymptote} up to third order in $\xi$, and the vertical markers, \ding{120}, on each dashed line show the 99\% validity range of this approximation, obtained from Eq.~\eqref{eq:validity-range}.}
  \label{fig:capcoupling-normalmodefreq-enlarged}
\end{figure}

The numerical values of the normal-mode frequencies as a function of the displacement parameter $\xi$, for different values of the capacitive coupling, are shown in Fig.~\ref{fig:capcoupling-normalmodefreq}. For the completely decoupled case (dotted curves), there is a degeneracy of mode frequency at points where the dotted curves intersect each other. On the other hand, when a capacitive coupling is present between two TL resonators, the degeneracy is lifted to give independent modes.

Figure~\ref{fig:capcoupling-normalmodefreq}(a) corresponds to the strong capacitive-coupling limit, where $C_\mathrm{c}\rightarrow \infty$ or $\omega_\mathrm{c}\rightarrow 0$. This corresponds to a perfectly transparent membrane inside a cavity where the displacement of the membrane has no effect on the mode structure. The curves attain more curvature as the capacitive coupling strength decreases ($C_\mathrm{c}\rightarrow 0$ or $\omega_\mathrm{c} \rightarrow\infty$) and, as in Fig.~\ref{fig:capcoupling-normalmodefreq}(d), eventually approach the dotted curves (decoupled case). 

It is clearly seen in Fig.~\ref{fig:capcoupling-normalmodefreq-enlarged} that Eq.~\eqref{eq:normalmodefreq-asymptote} fits well with the numerical values in Fig.~\ref{fig:capcoupling-normalmodefreq} in the vicinity of $\xi=0$. The range of $\xi$ where this approximation is valid varies between different coupling strengths. If the capacitive coupling is weak, the second-order coefficient $\omega_n^{(2)}$ has a large absolute value, which results in a stronger dependence of the normal-mode frequencies on $\xi$. At the same time, the range of $\xi$ where the approximation holds becomes shorter. For a strong capacitive coupling, however, the normal-mode frequencies are less sensitive to variations in $\xi$, with small expansion coefficients, and the validity range for the approximation is longer.

Following the normalization procedure using Sturm-Liouville theory of differential equations, which for example is employed in Refs.~\cite{Bourassa:2009gy, Bourassa:2012ej}, it is possible to express the mode function $u_n(x)$ as follows:
\begin{align}
u_n(x) &= N_n\bigg\{\Theta(-x)\frac{\sin{[k_n(x+d_\mathrm{L})]}}{\cos{(k_nd_\mathrm{L})}} \notag\\
	&\quad\qquad+\Theta(x)\frac{\sin{[k_n(x-d_\mathrm{R})]}}{\cos{(k_nd_\mathrm{R})}}\bigg\},
\end{align}
where
\begin{align}
N_n =\left[\frac{2\left(1+\frac{v_0}{\omega_\mathrm{c} d}\right)}{\frac{d_\mathrm{L}}{d}\sec{^2 (k_nd_\mathrm{L})}+\frac{d_\mathrm{R}}{d}\sec{^2(k_nd_\mathrm{R})}+\frac{\omega_\mathrm{c}}{k_n^2v_0d}}\right]^{1/2}
\end{align}
are the normalization constants chosen to satisfy
\begin{align}
c_0\int_{-d_\mathrm{L}}^{d_\mathrm{R}} u_n(x) u_m(x)\: \mathrm{d} x + C_\mathrm{c} (\Delta u_m)(\Delta u_n) &= C_{\Sigma}\delta_{nm},\notag\\
\frac{1}{\ell_0} \int_{-d_\mathrm{L}}^{d_\mathrm{R}} u_n'(x) u_m'(x)\: \mathrm{d} x &= \frac{1}{L_m} \delta_{nm}.
\end{align}
Here, $(\Delta u_m) \equiv u_m(0^+)-u_m(0^-)$ is the discontinuity of the mode functions at $x=0$, $C_{\Sigma}\equiv c_0 d + C_\mathrm{c}$ is the total capacitance of the system, and $L_m\equiv (\omega_m^2 C_\Sigma)^{-1}$ is the effective inductance for different modes \cite{Bourassa:2012ej}. The flux can be expressed in terms of the mode functions as $\Phi(x,t) = \sum_{n=0}^{\infty} u_n(x)\psi_n(x)$.

\begin{figure}
	\includegraphics[width=0.49\textwidth]{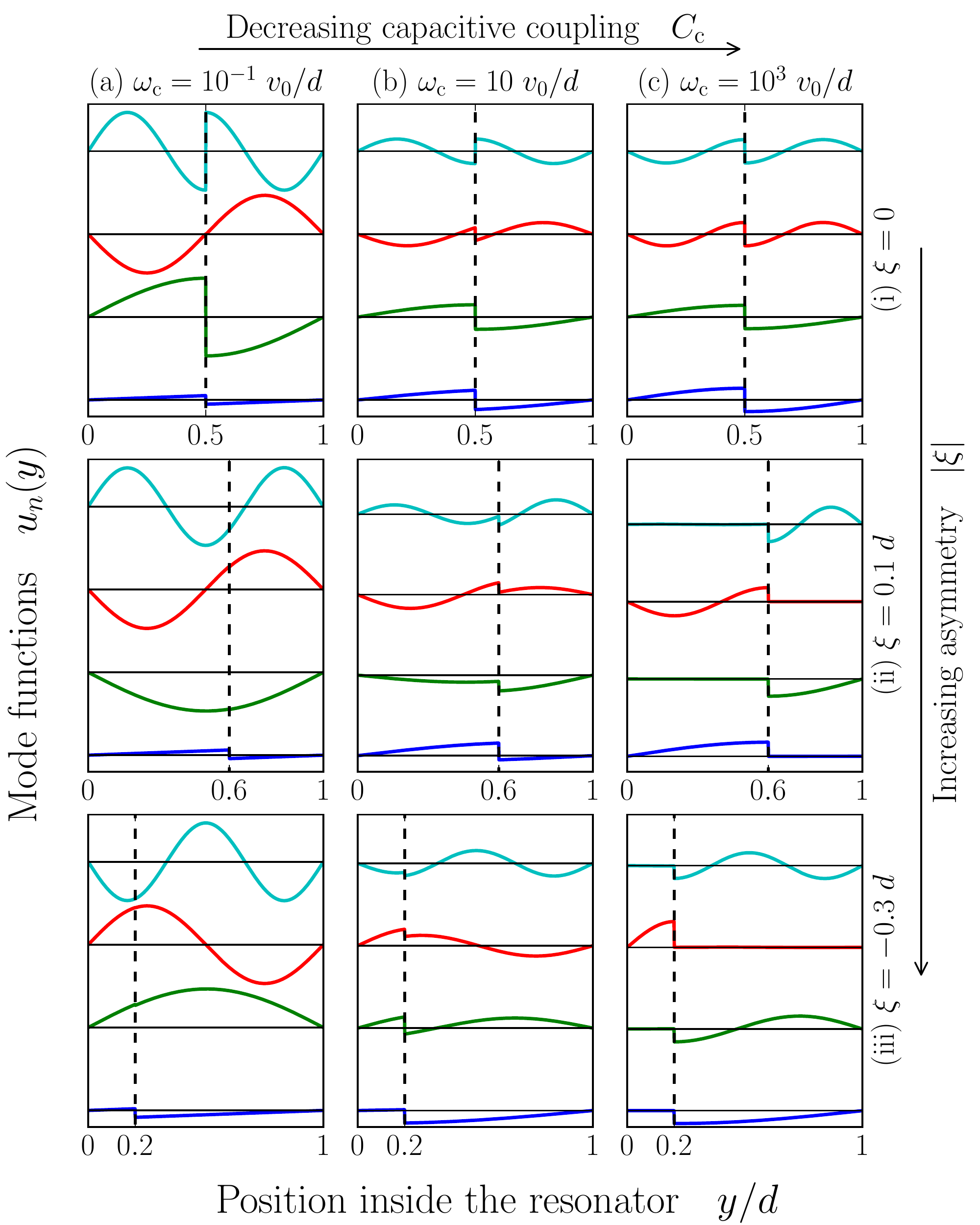}
	\caption{(color online) Normal-mode functions of two capacitively-coupled TL resonators as a function of the position $y$ inside the resonator for characteristic frequencies of the capacitive coupling: (a) $\omega_\mathrm{c} = 10^{-1}\ v_0/d$, (b) $\omega_\mathrm{c} = 10\ v_0/d$, (c) $\omega_\mathrm{c} = 10^{3}\ v_0/d$, and displacements of the capacitor: (i) $\xi= 0$, (ii) $\xi=0.1\ d$, (iii) $\xi = -0.3\ d$. The capacitive coupling is decreased from (a) to (c), and the asymmetry is increased from (i) to (iii). In each panel, the four curves represent the first four normal-modes ($n=0,1,2,3$) of the system from bottom to top. For clarity, the vertical axes are displaced for different modes. The coordinate describing the position in the resonator is shifted with $y=x+d_\mathrm{L}$ in such a way that $y=0$ and $y=d$ correspond to both ends of the resonator, i.e., $x=-d_\mathrm{L}$ and $x=d_\mathrm{R}$, respectively. The position of the capacitor $y=d/2+\xi$ is marked with vertical dashed lines.}
	\label{fig:capcoupling-mode-function}
\end{figure}

Figure~\ref{fig:capcoupling-mode-function} shows the mode functions for the few lowest modes. For the perfectly symmetric case [Fig.~\ref{fig:capcoupling-mode-function}(i), $\xi=0$], two nearby modes ($n=0$ and $n=1$, for instance) approach each other as the capacitive coupling decreases, and coalesce into a single mode in the end. In general, the $(2n)$-th and $(2n+1)$-th global mode of the system condense into one forming a twofold degeneracy (represented as intersections between dashed curves in Fig.~\ref{fig:capcoupling-normalmodefreq}), and these degenerate modes correspond to the local uncoupled modes for both TL resonators.

For the asymmetric case [Fig.~\ref{fig:capcoupling-mode-function}(ii, iii), $\xi\neq0$], as the capacitive coupling decreases, a global mode of the system reduces into a local uncoupled mode of either one of the two TL resonators; the spatial mode function is non-zero for one TL resonator and zero for the other TL resonator. A global mode reduces to a local uncoupled TL resonator mode with the closest mode frequency ($\omega_k^\alpha = \frac{2\pi v_0}{d_\alpha}(k+\frac{1}{2})$, $k=0, 1, 2, \ldots$).

Our discussion on capacitively-coupled TL resonators lead to the possibility of using electrical circuit elements to realize an optical cavity with a semi-transparent membrane inside.
\subsection{Tunable resonator}\label{sec:squidres}
In this section, we look into the mode structure of a SQUID-terminated TL resonator, which will be termed as a \emph{tunable resonator}. This system has been used in the realization of the DCE \cite{Johansson:2009zz, Johansson:2010bx, Wilson:2011wr, Johansson:2013tk} and in a circuit-analog of linear optomechanics \cite{johansson:2014}.
\begin{figure}
	\includegraphics[width=0.5\textwidth]{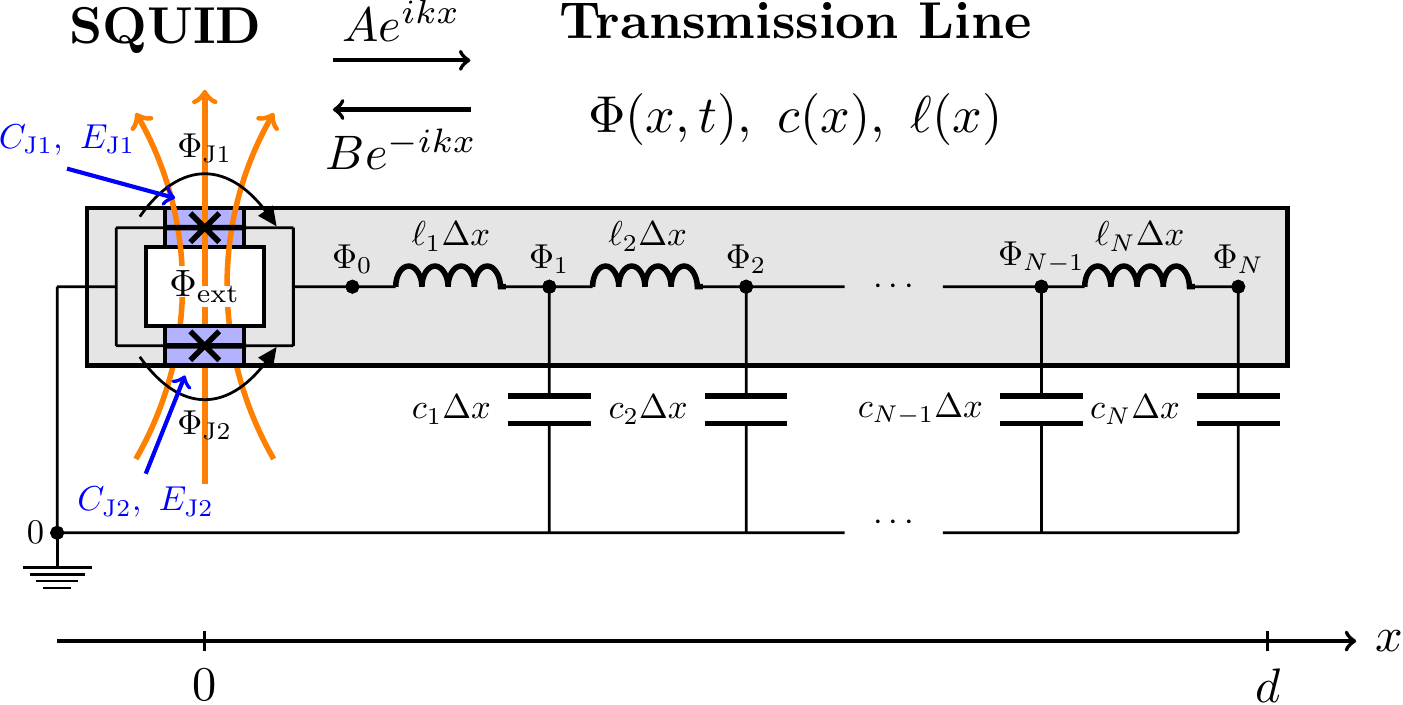}
	\caption{(color online) SQUID-terminated TL resonator expressed as a lumped-element circuit. The SQUID consists of two Josephson junctions on a loop through which an external flux $\Phi_\textrm{ext}$ is applied. Each junction in the SQUID has the capacitance and the Josephson energy $C_{\mathrm{J}\beta}$ and $E_{\mathrm{J}\beta}$ ($\beta = 1, 2$), respectively. The flux across each junction is denoted as $\Phi_{\mathrm{J}1}$ and $\Phi_{\mathrm{J}2}$. One side of the SQUID is grounded, and the opposite side is connected to a TL resonator with characteristic capacitance per unit length $c(x)$ and inductance per unit length $\ell(x)$ in the continuum limit. $\Phi(x,t)$ is the flux of the TL resonator at position $x$ and time $t$.}
	\label{fig:squidres-schematic}
\end{figure}

We consider the configuration described in Fig.~\ref{fig:squidres-schematic}. The fluxes across the Josephson junctions, $\Phi_{\mathrm{J}1}$ and $\Phi_{\mathrm{J}2}$, and the flux threading the SQUID loop $\Phi_\mathrm{ext}$ satisfy the fluxoid quantization relation \cite{tinkham},
\begin{align}
\Phi_{\mathrm{J}1} - \Phi_{\mathrm{J}2} = \Phi_\mathrm{ext}\quad (\textrm{mod }\Phi_0), \label{eq:fluxoid-quant}
\end{align}
where $\Phi_0\equiv h/2e$ is the magnetic flux quantum. We assume a symmetric SQUID configuration, with $E_{\mathrm{J}\beta}=E_{\mathrm{J}0}$ and $C_{\mathrm{J}\beta} = C_{\mathrm{J}}/2$ ($\beta=1, 2$). In this case, the SQUID behaves like a single Josephson junction with effective capacitance $C_{\mathrm{J}}$ and flux-dependent Josephson energy,
\begin{align}
E_\mathrm{J}(\Phi_\mathrm{ext})=2E_{\mathrm{J}0}\left\vert \cos{\left(\pi\frac{\Phi_\mathrm{ext}}{\Phi_0}\right)}\right\vert.
\end{align}
We define $\Phi_{\mathrm{J}}\equiv(\Phi_{\mathrm{J}1}+\Phi_{\mathrm{J}2})/2$ as the flux across the SQUID. In our system, this flux is related to the flux of the TL resonator $\Phi(x,t)$ as $\Phi_\mathrm{J} = -\Phi(0,t)$. With these parameters, the Lagrangian density $\mathcal{L}$ of the system is given by the following \cite{yurke:1984, devoret:1995}:
\begin{align*}
\mathcal{L} &= \frac{c(x)}{2}\left[\partial_t \Phi(x,t)\right]^2 - \frac{1}{2\ell(x)}\left[\partial_x \Phi(x,t)\right]^2 \\
		&\ + \left\{ \frac{C_\mathrm{J}}{2} \left[\partial_t\Phi(x,t)\right]^2 + E_\mathrm{J}(\Phi_\mathrm{ext})\cos{\left[2\pi\frac{\Phi(x,t)}{\Phi_0}\right]}\right\}\delta(x),
\end{align*}
$(0<x<d)$. Here, $c(x)$ and $\ell(x)$ are the characteristic capacitance and inductance per unit length of the TL resonator, respectively. The Euler-Lagrange equation of motion Eq.~\eqref{eq:euler-lagrange} yields the partial differential equation Eq.~\eqref{eq:partial-diff-tl} for $x>0$, as in Sec.~\ref{sec:capcoupling}. The boundary condition at $x=0$ is given by:
\begin{align*}
0 & =  C_\mathrm{J} \partial_{tt}\Phi(0,t) -\frac{1}{\ell(0)}\partial_x \Phi(0, t) \\
   &\quad + \left(\frac{2\pi}{\Phi_0}\right)E_\mathrm{J}(\Phi_\textrm{ext})\sin{\left[2\pi \frac{\Phi(0,t)}{\Phi_0}\right]}.
\end{align*}
We consider the ground-ended TL resonator.
Assuming that the TL resonator is uniform, i.e., $c(x) = c_0$ and $\ell(x) = \ell_0$, our problem reduces to solving the massless Klein-Gordon wave equation Eq.~\eqref{eq:wave-eqn} for $x>0$, subject to the boundary conditions,
\begin{subequations}
\begin{align}
0 &= C_\mathrm{J} \partial_{tt}\Phi(0,t) -\frac{1}{\ell_0}\partial_x \Phi(0, t)+\frac{1}{L_\mathrm{J}}\Phi(0,t),\label{eq:squidres-bd1}\\
0&= \Phi(d,t).\label{eq:squidres-bd2}
\end{align}
\end{subequations}
Here, we have made an assumption that the phase across the SQUID is small, $2\pi\Phi(0,t)/\Phi_0 \ll 1$, and expanded the sine function to second order in $2\pi\Phi(0,t)/\Phi_0$. This amounts to replacing the cosine potential with an effective inductor \cite{Bourassa:2012ej}, whose inductance,
\begin{align}
L_\mathrm{J}\equiv \frac{1}{E_\mathrm{J}(\Phi_\mathrm{ext})}\left(\frac{\Phi_0}{2\pi}\right)^2 = \frac{1}{2E_{\mathrm{J}0}\left|\cos{\left(\pi\frac{\Phi_\mathrm{ext}}{\Phi_0}\right)} \right|}\left(\frac{\Phi_0}{2\pi}\right)^2,
\end{align}
can be adjusted by the external flux. We also define $L_{\mathrm{J}0}$ as the effective inductance $L_\mathrm{J}$ for zero external flux. i.e., $L_\mathrm{J} = L_{\mathrm{J}0}|\sec{(\pi\Phi_\mathrm{ext}/\Phi_0)}|$. Let us assume for now that the external flux $\Phi_\mathrm{ext}$ is constant in time and leave out the flux dependence for notational convenience.

The separation of variables $\Phi(x,t)=u(x)\psi(t)$ gives two ordinary differential equations Eq.~\eqref{eq:two-ode} with a constant $k=\omega/v_0$ ($v_0\equiv1/\sqrt{\ell_0c_0}$).  The general solution for $u(x)$ is given by
\begin{align}
u(x) = A e^{ikx} + Be^{-ikx}\quad (0<x<d).
\end{align}
The normal-mode frequencies of the system are determined by Eqs.~\eqref{eq:squidres-bd1}-\eqref{eq:squidres-bd2}. The equation for the normal-mode frequency is given by
\begin{align}
\tan{\left(\frac{\omega_n d}{v_0}\right)} = - \frac{\omega_n /v_0}{1-(\omega_n/\omega_\mathrm{J})^2}\frac{L_\mathrm{J}}{\ell_0},\label{eq:squidres-eigenmode-1}
\end{align}
where $n$ is used to label the discrete modes. Here, $\omega_\mathrm{J} \equiv 1/\sqrt{C_\mathrm{J} L_\mathrm{J}}$ is the plasma frequency of the SQUID. We define the ratio of the mode frequency of the system to the plasma frequency of the SQUID as $\eta_n \equiv\omega_n/\omega_\mathrm{J}$.

If we only excite modes which oscillate much slower than the plasma frequency of the SQUID, i.e., $\eta_n\rightarrow0$, then Eq.~\eqref{eq:squidres-eigenmode-1} can be written as
\begin{subequations}
\begin{align}
\tan{(k_nd)} = -k_n\Delta d,\label{eq:squidres-eigenmode-2}
\end{align}
where $k_n = \omega_n / v_0$ and $\Delta d=L_\mathrm{J}/\ell_0$ is the length of the TL resonator whose total inductance equals to the effective SQUID inductance $L_\mathrm{J}$. If we further assume that this length is short compared to the mode wavelength, $\epsilon \sim k\Delta d\ll1$, Eq.~\eqref{eq:squidres-eigenmode-2} can be rewritten as
\begin{align}
\tan{\left[k_n(d+\Delta d)\right]} = \mathcal{O}(\epsilon^3).\label{eq:squidres-eigenmode-3}
\end{align}
\end{subequations}
The analytic expression for the mode frequencies obtained from Eq.~\eqref{eq:squidres-eigenmode-3} are
\begin{align}
k_n = \frac{n\pi}{d+\Delta d} \quad(n=1,2,3,\ldots).\label{eq:squidres-normalmode}
\end{align}
This means that, up to second order in $\epsilon$, $\Delta d$ can be interpreted as an additional \emph{effective length} of the TL resonator introduced by the SQUID. This effective length can be tuned with the external flux $\Phi_\mathrm{ext}$.

\begin{figure}
  \includegraphics[width=.48\textwidth]{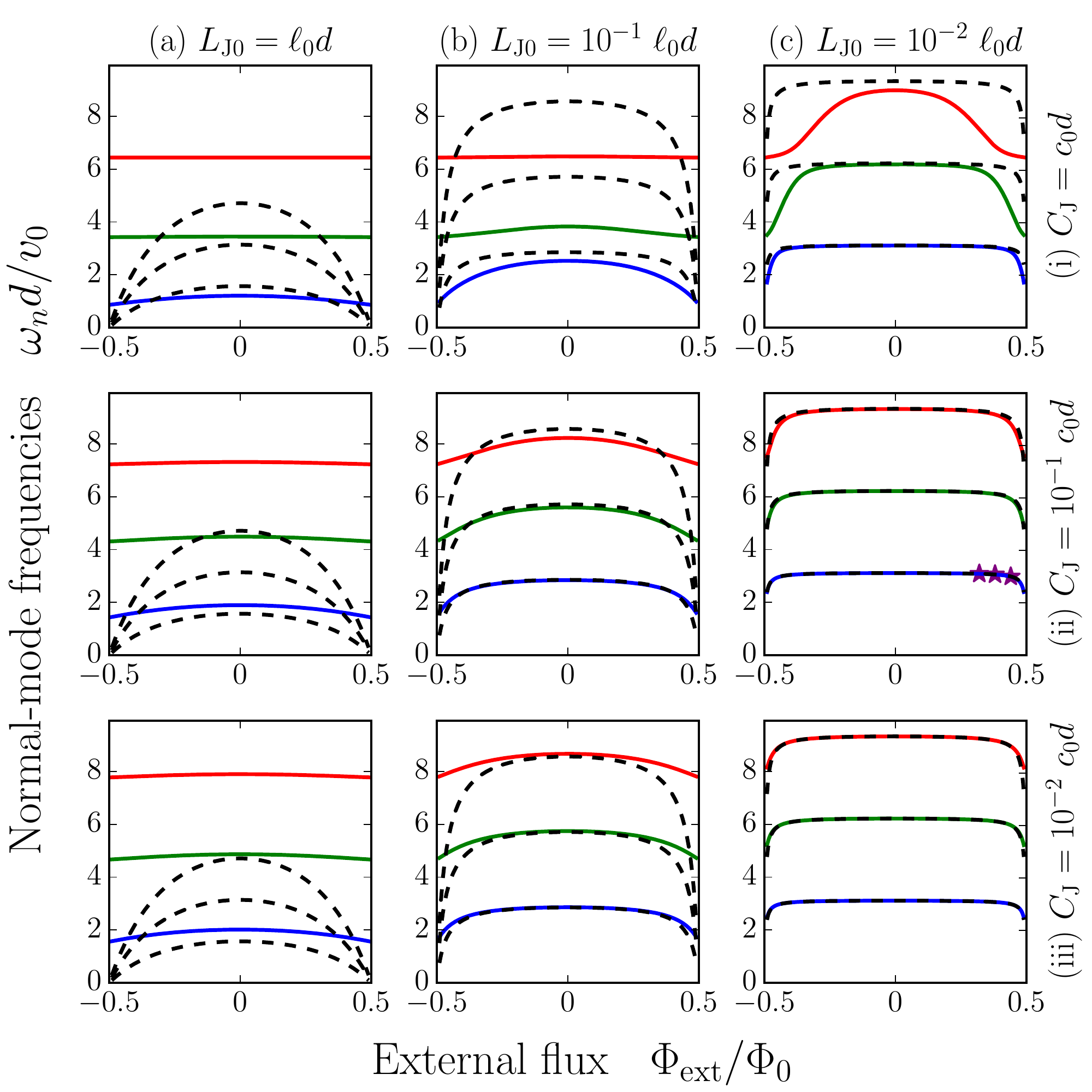}
  \caption{(color online) Normal-mode frequencies of the SQUID-terminated resonator as a function of the external flux $\Phi_\mathrm{ext}$, for (a) $L_{\mathrm{J}0} = \ell_0 d$, (b) $L_{\mathrm{J}0} = 10^{-1}\ \ell_0 d$, (c) $L_{\mathrm{J}0} = 10^{-2}\ \ell_0 d$ and (i) $C_{\mathrm{J}} =  c_0 d$, (ii) $C_{\mathrm{J}} = 10^{-1}\ c_0 d$, (iii) $C_{\mathrm{J}} = 10^{-2}\ c_0 d$. The three lines of each panel correspond to the first three modes ($n=1, 2, 3$), from bottom to top. The numerical values of the normal-mode frequencies obtained from Eq.~\eqref{eq:squidres-eigenmode-1} (solid), and the analytical result from Eq.~\eqref{eq:squidres-normalmode} (dashed) are compared. The three markers, {\ding{72}}, in the [(c), (ii)] panel, which are the fundamental-mode frequencies for $\Phi_\mathrm{ext}/\Phi_0 = \{0.32,\ 0.38,\ 0.44\}$, correspond to the mode functions depicted in Fig.~\ref{fig:squidres-normalmodefunc}.}
	\label{fig:squidres-normalmodefreq}
\end{figure}
\begin{figure}
  \includegraphics[width=.48\textwidth]{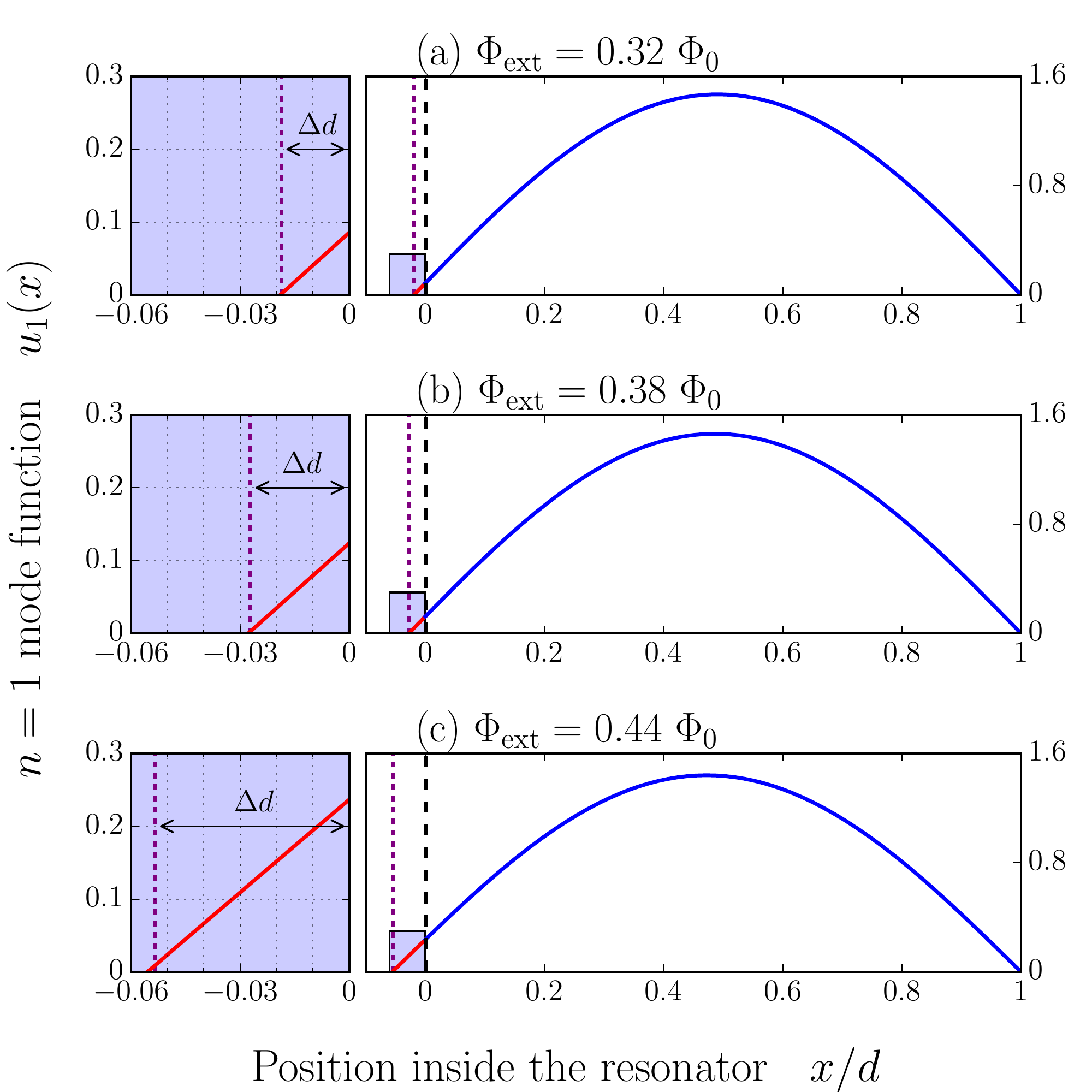}
  \caption{(color online) The $n=1$ mode functions of the SQUID-terminated resonator system, under the condition of $L_{\mathrm{J}0}=10^{-2}\ \ell_0d$, $C_\mathrm{J}=10^{-1}\ c_0d$, and (a) $\Phi_\mathrm{ext}=0.32\ \Phi_0$, (b) $\Phi_\mathrm{ext}=0.38\ \Phi_0$, (c) $\Phi_\mathrm{ext}=0.44\ \Phi_0$ (marked with {\ding{72}} in Fig.~\ref{fig:squidres-normalmodefreq}[(c), (ii)]). The real mode functions ($x>0$, blue) are distinguished from the virtual mode functions ($x<0$, red) by the real end ($x=0$, dashed) of the SQUID-terminated resonator. The effective length $\Delta d$ obtained from Eq.~\eqref{eq:squidres-eigenmode-2} are marked with vertical dotted lines. The panels on the left correspond to the shaded areas of the panels on the right. It is seen that the $x$-intercept of each panel is in good agreement with $\Delta d$.}
	\label{fig:squidres-normalmodefunc}
\end{figure}

Figure~\ref{fig:squidres-normalmodefreq} shows the comparison of the numerical result of Eq.~\eqref{eq:squidres-eigenmode-1} and the analytical expression from the effective length interpretation, Eq.~\eqref{eq:squidres-normalmode}, for several orders of magnitude of $C_\mathrm{J}$ and $L_{\mathrm{J}0}$. The numerical values of the normal-mode frequencies shows a substantial deviation from the analytical result for large values of $L_{\mathrm{J}0}$ and $C_\mathrm{J}$. This is due to the fact that the plasma frequency of the SQUID decreases for larger values of the effective inductance $L_{\mathrm{J}0}$ and capacitance $C_\mathrm{J}$, which undermines our assumption that $\eta_n\rightarrow 0$. In general, the discrepancy between the numerical and the analytical result is larger for higher $n$ modes, and for values of $\Phi_\mathrm{ext}$ closer to half-integer multiples of a flux quantum (e.g., $\pm0.5\ \Phi_0$, $\pm1.5\ \Phi_0$, etc.). Thus, it is safe to use low values of $L_{\mathrm{J}0}$ and $C_\mathrm{J}$ in order to use Eq.~\eqref{eq:squidres-normalmode} in our discussion.

However, there is a disadvantage of using too small values of $C_\mathrm{J}$ and $L_{\mathrm{J}0}$: the normal-mode frequencies become insensitive to variations in the external flux, as can be seen in Fig.~\ref{fig:squidres-normalmodefreq}. That is, the validity of the analytical expression comes at the expense of the tunability of the system. Thus, it is important that we find the optimal range of $L_{\mathrm{J}0}$ and $C_\mathrm{J}$, suitable to specific cases. If the external flux is not too close to half-integral multiples of $\Phi_0$, the numerical values agree well with the analytical expression as long as $C_\mathrm{J} \le 10^{-1}\ c_0 d$, and $L_{\mathrm{J}0} \le 10^{-2}\  \ell_0 d$. In this regime, the analytical expression Eq.~\eqref{eq:squidres-normalmode} is valid.

The normal-mode functions of the system $u_n(x)$ are given by
\begin{align}
u_n(x) = N_n \frac{\sin{[k_n(x-d)]}}{\cos{(k_nd)}}\quad (0<x<d),
\end{align}
where
\begin{align*}
N_n = \left[ \frac{2\left(1+\frac{C_\mathrm{J}}{c_0 d}\right)}{\sec{^2(k_nd)} + \frac{L_\mathrm{J}}{\ell_0 d} \frac{1+\eta_n^2}{(1-\eta_n^2)^2}} \right]^{1/2}
\end{align*}
are the normalization constants chosen to satisfy
\begin{align*}
c_0\int_{0}^{d} u_n(x) u_m(x)\: \mathrm{d} x + C_\mathrm{J} u_m(0) u_n(0) &= C_{\Sigma}\delta_{nm},\notag\\
\frac{1}{\ell_0} \int_{0}^{d} u_n'(x) u_m'(x)\: \mathrm{d} x + \frac{1}{L_\mathrm{J}} u_m(0) u_n(0)&= \frac{1}{L_m} \delta_{nm}.
\end{align*}
Here, the total capacitance of the system $C_\Sigma=c_0d+C_\mathrm{J}$, and the effective mode inductances $L_m\equiv (\omega_m^2 C_\Sigma)^{-1}$ are defined in the same way as in Sec.~\ref{sec:capcoupling}.

The fundamental mode function, $u_1(x)$, for certain values of $\Phi_\mathrm{ext}$ is illustrated in Fig.~\ref{fig:squidres-normalmodefunc}. Here, the mode function is zero at the end without the SQUID ($x=d$), but is non-zero at the other end with the SQUID ($x=0$). If we continuously extend the mode function to $x<0$, an $x$-intercept takes place. This point, which arise from the shift of mode frequencies due to the presence of the SQUID, can be interpreted as the virtual end of the TL resonator.

The distance between the real end ($x=0$) and the virtual end ($x$-intercept) can be interpreted as an additional virtual length of the resonator. Note that this definition of virtual length in Fig.~\ref{fig:squidres-normalmodefunc} is in accordance with the effective length $\Delta d$ in Eq.~\eqref{eq:squidres-eigenmode-2}, which is defined as the effective inductance of the SQUID divided by the characteristic inductance per unit length of the TL resonator. The virtual length becomes longer if we increase the external flux; it becomes shorter as we decrease the external flux. Also, under small variations in $\Phi_\mathrm{ext}$, this change in virtual length can be approximated as linear \cite{johansson:2014}:
\begin{align}
\Delta d (\Phi_\mathrm{ext}^0+\delta\Phi_\mathrm{ext}) \approx \Delta d^{(0)} + \Delta d^{(1)} \delta\Phi_\mathrm{ext},\label{eq:effective-length-linear}
\end{align}
with the expansion coefficients given by
\begin{align}
\begin{split}
\Delta d^{(0)} &= \left(\frac{\Phi_0}{2\pi}\right)^2 \frac{1}{\ell_0 E_{\mathrm{J}0}},\\
\Delta d^{(1)} &= \frac{1}{2}\left(\frac{\Phi_0}{2\pi}\right) \frac{1}{\ell_0 E_{\mathrm{J}0}} \tan{\left(\pi\frac{\Phi_\mathrm{ext}^0}{\Phi_0}\right)}.
\end{split} \label{eq:effective-length-linear-coeff}
\end{align}
Therefore, under valid assumptions, the SQUID-terminated TL resonator can be thought of as a cavity whose total length can be linearly tuned with the external flux. Hereafter, we call this configuration a tunable resonator.

\subsection{Capacitively-coupled tunable resonators}\label{sec:analogcircuit}
\begin{figure}
	\includegraphics[width=0.48\textwidth]{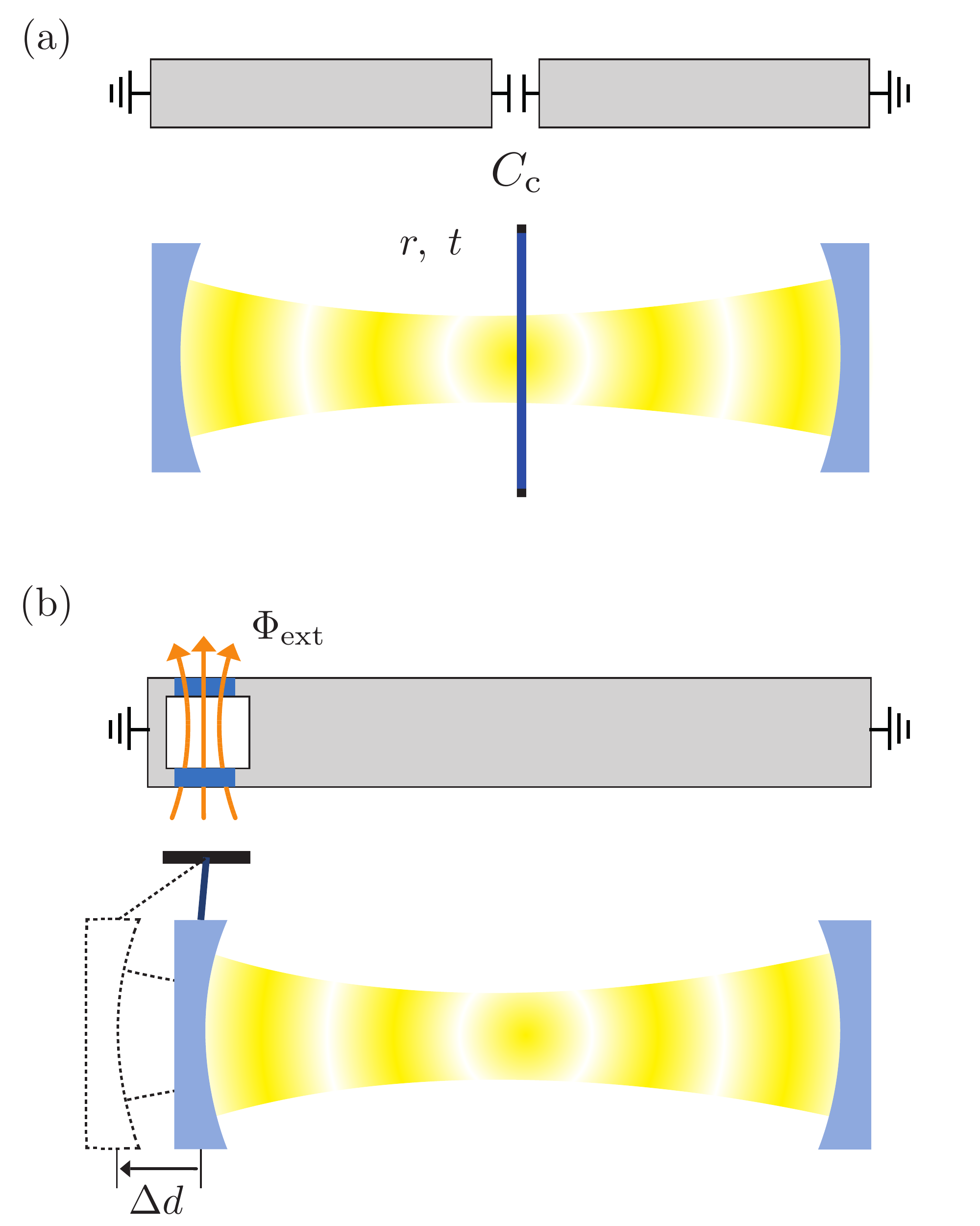}
	\caption{(color online) Schematic diagram of the analogy of (a) a semi-transparent membrane and (b) a movable mirror in electrical circuits.}
	\label{fig:analogcircuit-analogy-1}
\end{figure}
\begin{figure}
	\includegraphics[width=0.48\textwidth]{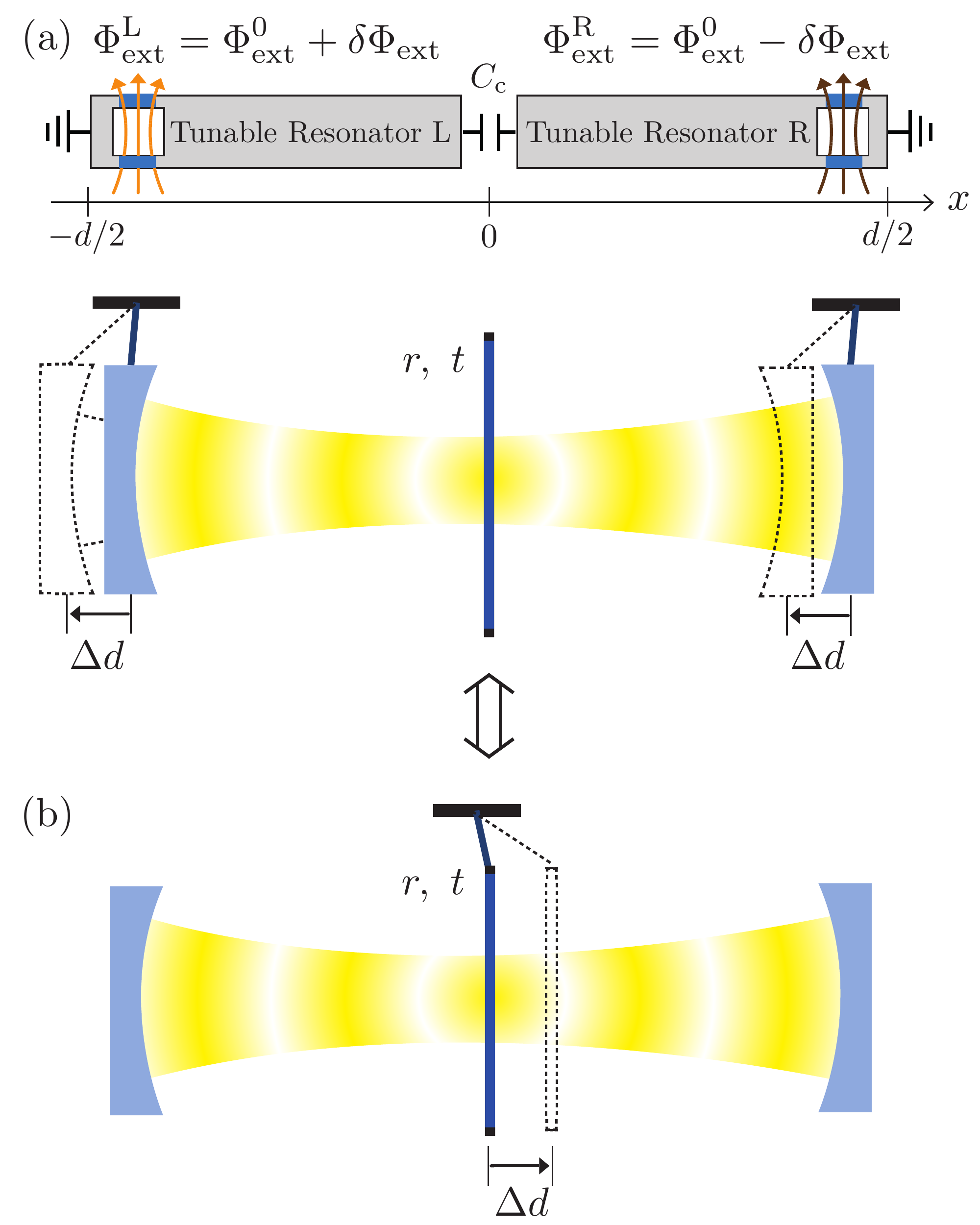}
	\caption{(color online) Combining the principles shown in Fig.~\ref{fig:analogcircuit-analogy-1}, an analog circuit [(a), up] of the system of [(a), down] can be designed to give the quadratic coupling of optomechanics. The motion of the virtual ends of the tunable resonators are synchronized so as to maintain the total effective length of the resonator unchanged. Therefore, in the co-moving frame with the effective cavity, (b) is equivalent to a cavity consisting of fixed mirrors with a semi-transparent membrane moving inside [(a), down].}
	\label{fig:analogcircuit-analogy-2}
\end{figure}

Now that we have a semi-transparent membrane (optics) and a movable mirror (mechanics) for electrical circuits, we move to the discussion of combining these elements to generate the desired coupling of mechanical and optical degrees of freedom. The outline is described in Figs.~\ref{fig:analogcircuit-analogy-1}-\ref{fig:analogcircuit-analogy-2}.

From the analogy illustrated in Fig.~\ref{fig:analogcircuit-analogy-1}, it is natural to think of capacitively-coupled tunable resonators, which look like Fig.~\ref{fig:analogcircuit-analogy-2}[(a), up], in realizing Fig.~\ref{fig:analogcircuit-analogy-2}(b). In Fig.~\ref{fig:analogcircuit-analogy-2}(a), two homogeneous and uniform SQUID-terminated TL resonators are capacitively coupled to each other with a capacitor $C_\mathrm{c}$ in between at $x=0$.

Following the convention of the previous sections, we assume that the characteristic capacitance and inductance per unit length of both TL resonators are $c_0$ and $\ell_0$, and that all the Josephson junctions have equal capacitance $C_\mathrm{J}/2$ and the Josephson energy $E_{\mathrm{J}0}$. Also, we assume that the total length of the system is $d$, with each tunable resonator ranging over $(-\frac{d}{2},0)$ and $(0,\frac{d}{2})$.

Motivated by Fig.~\ref{fig:analogcircuit-analogy-2}, it is expected that the tunable resonators L and R can be considered as one-sided cavities of effective lengths $d_\mathrm{L} = d/2 + \Delta d_\mathrm{L}$ and $d_\mathrm{R} = d/2 + \Delta d_\mathrm{R}$, where $\Delta d_\alpha$ is the additional effective length, Eq.~\eqref{eq:squidres-eigenmode-2}, arising from the flux threading each SQUID, $\Phi_\mathrm{ext}^\alpha$ ($\alpha=\mathrm{L},\ \mathrm{R}$). Also, the capacitive coupling should operate as a semi-transparent optical membrane connecting two one-sided cavities of effective lengths $d_\mathrm{L}$ and $d_\mathrm{R}$.

The Lagrangian of the system can be written as 
\begin{align}
L = L_\mathrm{tl} + L_\mathrm{c}+L_\mathrm{s}^{\mathrm{L}}+L_\mathrm{s}^{\mathrm{R}}, \label{eq:analogy-lagrangian}
\end{align}
where $L_\mathrm{tl}$ is the Lagrangian of the TL resonators, $L_\mathrm{c}$ is the Lagrangian of the capacitor in the middle, and $L_\mathrm{s}^{\alpha}$ is the Lagrangian of the SQUID ($\alpha=\mathrm{L},\mathrm{R}$), each given by
\begin{align*}
L_\mathrm{tl}& = \int_{-d/2}^{d/2} \left\{\frac{c_0}{2}\left[\partial_t\Phi(x,t)\right]^2 - \frac{1}{2\ell_0} \left[\partial_x\Phi(x,t)\right]^2\right\}\mathrm{d} x,\\
L_\mathrm{c}& = \frac{C_\mathrm{c}}{2} \left[\partial_t\Phi(0^+,t) - \partial_t\Phi(0^-,t)\right]^2,\\
L_\mathrm{s}^\alpha& = \frac{C_\mathrm{J}}{2} \left[\partial_t\Phi\left(s_\alpha,t\right)\right]^2 -\frac{L_{\mathrm{J}}^\alpha}{2} \left[{\Phi(s_\alpha,t)}\right]^2,
\end{align*}
($s_\mathrm{L} = -d/2$, $s_\mathrm{R} = d/2$). Here, we assumed that the system is in the phase regime where the fluxes across the SQUIDs are small, $2\pi\Phi(s_\alpha,t)/\Phi_0\ll1$, and replaced the nonlinear potential with effective flux-dependent inductors with inductances
\begin{align*}
L_\mathrm{J}^\alpha \equiv \frac{1}{2E_{\mathrm{J}0}}\left(\frac{\Phi_0}{2\pi}\right)^2\left|\sec{\left(\pi\frac{\Phi_\mathrm{ext}^\alpha}{\Phi_0}\right)}\right|.
\end{align*}

It can be shown that detailed calculations using the Euler-Lagrange equation of motion and Sturm-Liouville theory of differential equations yield the intuitive result,
\begin{align}
\begin{split}
\frac{\omega_\mathrm{c}}{\omega_n} &=\tan{\left[\frac{\omega_n}{v_0} \left(\frac{d}{2}+\Delta d_\mathrm{L}\right)\right]}\\
& \quad+ \tan{\left[\frac{\omega_n}{v_0} \left(\frac{d}{2}+\Delta d_\mathrm{R}\right)\right]}+ \mathcal{O}(\epsilon^3),
\end{split}\label{eq:analogcircuit-eigenmode}
\end{align}
which is the eigenmode equation for capacitive coupling, Eq.~\eqref{eq:capcoupling-mode-1}, with effective cavity lengths on the sides given as $d_\mathrm{L} = d/2+\Delta d_\mathrm{L}$ and $d_\mathrm{R} = d/2 + \Delta d_\mathrm{R}$. Note that Eq.~\eqref{eq:analogcircuit-eigenmode} is obtained under the assumption that the mode frequency is much lower than the plasma frequency of the SQUID ($\eta_n^\alpha\rightarrow 0$). Also, the additional effective lengths of the SQUIDs are taken as small parameters, $\epsilon\sim k\Delta d_\mathrm{L},\ k\Delta d_\mathrm{R} \ll 1$. This equation makes it possible to expand the normal-mode frequencies with respect to the total effective length of the system,
$$D = d_\mathrm{L}+d_\mathrm{R} =d + \Delta d_\mathrm{L} +\Delta d_\mathrm{R},$$
and the difference in the effective lengths,
$$\xi = \frac{d_\mathrm{L} - d_\mathrm{R}}{2} = \frac{\Delta d_\mathrm{L} - \Delta d_\mathrm{R}}{2},$$
using Eq.~\eqref{eq:normalmodefreq-asymptote}.

In the configuration of Fig.~\ref{fig:analogcircuit-analogy-2}[(a), up], the fluxes $\Phi_\mathrm{ext}^\mathrm{L}$ and $\Phi_\mathrm{ext}^\mathrm{R}$ through the SQUIDs are set to have the same bias flux $\Phi_\mathrm{ext}^0$. On top of the equal-bias fluxes, a small variation of the same magnitude $\delta\Phi_\mathrm{ext}$ is added in the opposite direction, i.e.,
\begin{align}
\begin{split}
\Phi_\mathrm{ext}^\mathrm{L} &= \Phi_\mathrm{ext}^0 + \delta\Phi_\mathrm{ext},\\
\Phi_\mathrm{ext}^\mathrm{R} &= \Phi_\mathrm{ext}^0 - \delta\Phi_\mathrm{ext}.
\end{split}
\end{align}
This results in a simultaneous movement of the virtual ends in the same direction. Here, the magnitude of the variation $|\delta\Phi_\mathrm{ext}|$ should be small enough compared to the magnetic flux quantum $\Phi_0$ to ensure that the effective lengths of the TL resonators, $\Delta d_\mathrm{L}$ and $\Delta d_\mathrm{R}$, change linearly with the flux displacement. In this regime, Eq.~\eqref{eq:effective-length-linear} is applicable and the additional effective length of each tunable resonator can be written as:
\begin{align}
\begin{split}
\Delta d_\mathrm{L} &= \Delta d^{(0)} + \Delta d^{(1)} \delta\Phi_\mathrm{ext},\\
\Delta d_\mathrm{R} &= \Delta d^{(0)} - \Delta d^{(1)} \delta\Phi_\mathrm{ext},
\end{split}
\end{align}
with the expansion coefficients Eq.~\eqref{eq:effective-length-linear-coeff}. Now, the total effective length of the system is a constant,
\begin{align}
D = d + \Delta d_\mathrm{L} + \Delta d_\mathrm{R}
	= d + 2\Delta d^{(0)},
\end{align}
and the displacement parameter is linear in the flux variation,
\begin{align}
\xi =\frac{\Delta d_\mathrm{L} - \Delta d_\mathrm{R}}{2} = \Delta d^{(1)} \delta\Phi_\mathrm{ext}.
\end{align}
Thus, up to third order in $\delta\Phi_\mathrm{ext}$, the normal-mode frequency becomes ($n=0, 1, 2, \ldots$)
\begin{align}
\omega_n \approx \omega_n^{(0)} \left[1-(-1)^n \frac{\omega_\mathrm{c}\left(\Delta d^{(1)}\right)^2}{ v_0 D}  {\delta\Phi_\mathrm{ext}^2}\right],\label{eq:mode-freq-optomechanics}
\end{align}
where the overall constant is given by
\begin{align}
\begin{split}
\omega_n^{(0)} &= \frac{\pi v_0}{D}\bigg( n+\textrm{mod}(n+1,2)\bigg) \\
			&\quad -\frac{2v_0\cos{^{-1}(|r_n^{(0)}|)}}{D}\textrm{mod}(n+1,2).
\end{split}\label{eq:unperturbed-normal-mode-freq}
\end{align}
Here, $|r_n^{(0)}|$ is the absolute value of the effective reflectivity corresponding to the mode $n$,
\begin{align}
|r_n^{(0)}| = \frac{{\omega_\mathrm{c}}/{2\omega_n^{(0)}}}{\sqrt{1+\left({\omega_\mathrm{c}}/{2\omega_n^{(0)}}\right)^2}}.\label{eq:reflectivity-absolute}
\end{align}
to zeroth order in $\delta\Phi_\mathrm{ext}$.

From the Sturm-Liouville theory of differential equation, the mode functions $u_n(x)$ ($n=0, 1, 2, \ldots$) should satisfy the orthonormality relation,
\begin{widetext}
\begin{align}
\begin{split}
c_0 \int_{-d/2}^{d/2} u_n(x)u_m(x)\: \mathrm{d} x + C_\mathrm{J} u_n\!\left(-\frac{d}{2}\right) u_m\!\left(-\frac{d}{2}\right)
+ C_\mathrm{c} (\Delta u_n) (\Delta u_m) + C_\mathrm{J} u_n\!\left(\frac{d}{2}\right) u_m\!\left(\frac{d}{2}\right)& = C_\Sigma \delta_{nm},\\
\frac{1}{\ell_0} \int_{-d/2}^{d/2} u_n'(x)u_m'(x)\: \mathrm{d} x + \frac{1}{L_{\mathrm{J}}^\mathrm{L}} u_n\!\left(-\frac{d}{2}\right) u_m\!\left(-\frac{d}{2}\right)
+ \frac{1}{L_\mathrm{J}^\mathrm{R}} u_n\!\left(\frac{d}{2}\right) u_m\!\left(\frac{d}{2}\right)& = \frac{1}{L_m} \delta_{nm}.
\end{split}
\end{align}
\end{widetext}
where $(\Delta u_n)\equiv u_n(0^+)-u_n(0^-)$ is the discontinuity of the mode function at $x=0$. Here, $C_{\Sigma} = c_0d + 2C_{\mathrm{J}} + C_\mathrm{c}$ is the total capacitance of the system and $L_m = (\omega_m^2 C_\Sigma)^{-1}$ are the effective inductances for different modes.

With this normalization, the flux can be expressed in terms of the mode functions as $\Phi(x,t) = \sum_{n=0}^{\infty} u_n(x)\psi_n(t)$. Plugging this into the Eq.~\eqref{eq:analogy-lagrangian}, the Lagrangian of the system can be simplified as
\begin{align}
L = L(\psi_j, \dot{\psi}_j; t) = \sum_{n=0}^{\infty} \left[ \frac{C_\Sigma}{2} \dot{\psi}_n^2 - \frac{\psi_n^2}{2L_n}\right].\label{eq:lagrangian-generalized-coord}
\end{align}
Here, the Lagrangian of the system is expressed with the mode fluxes $\psi_n(t)$ as generalized coordinates.

\section{Hamiltonian formulation}\label{sec:hamiltonian}
Now, we are able to simulate a semi-transparent membrane in an optical cavity with capacitively-coupled tunable resonators. The effective displacement of the membrane can be adjusted linearly with the variation in the external flux $\delta\Phi_\mathrm{ext}$.

In this section, we continue our discussion on the analog system of Sec.~\ref{sec:analogcircuit}, but now using a Hamiltonian formulation. In Sec.~\ref{sec:classical-opto}, we employ the canonical quantization procedure \cite{yurke:1984, devoret:1995} to derive the Hamiltonian of the classical quadratic optomechanical system, where the pseudo-mechanical degree of freedom (the variation in the external flux, $\delta\Phi_\mathrm{ext}$) remains classical. Furthermore, we introduce an additional quantum field to the external flux variation in Sec.~\ref{sec:quantum-opto}. This results in the quantum quadratic optomechanical coupling, where the pseudo-mechanical degree of freedom is quantum mechanical.

\subsection{Classical quadratic optomechanics}\label{sec:classical-opto}
We start from the Lagrangian of Eq.~\eqref{eq:lagrangian-generalized-coord}. The momentum $\theta_n$ conjugate to the mode flux $\psi_n$ is given by
\begin{align}
\theta_n = \frac{\partial L}{\partial \dot{\psi}_n} = C_{\Sigma} \dot{\psi}_n.
\end{align}
The Hamiltonian $H(\psi_j, \theta_j; t)$ is generated by the Legendre transformation \cite{goldstein}:
\begin{align}
H(\psi_j, \theta_j; t) &= \sum_{n=0}^{\infty} \dot{\psi}_{n}\theta_n - L(\psi_j, \dot{\psi}_j; t)\notag \\
				&= \sum_{n=0}^{\infty}\left[\frac{\theta_n^2}{2C_\Sigma} + \frac{C_\Sigma }{2}\omega_n^2\psi_n^2\right].
\end{align}
From now on, we treat the canonical variables $(\psi_n, \theta_n)$ as quantum operators that satisfy the canonical commutation relation \cite{yurke:1984, devoret:1995}:
\begin{align}
[\hat{\psi}_n, \hat{\theta}_m ] &= i\hbar\delta_{nm}.
\end{align}
This is equivalent to introducing the annihilation and the creation operators, $\hat{a}_n$ and $\hat{a}_n^\dagger$, with
\begin{align}
\begin{split}
\hat{\psi}_n &= \sqrt{\frac{\hbar}{2\omega_nC_\Sigma}} (\hat{a}_n^\dagger + \hat{a}_n),\\
\hat{\theta}_n &= i\sqrt{\frac{\hbar\omega_nC_\Sigma}{2}} (\hat{a}_n^\dagger - \hat{a}_n).
\end{split}
\end{align}
Note that the annihilation and the creation operators here are defined for the global mode, not for an individual tunable resonator forming the system. The annihilation operator $\hat{a}_n$ destroys one microwave photon of frequency $\omega_n$, from the system (i.e., removing a photon from the $n$-th global mode of the capacitively-coupled tunable resonators). The creation operator $\hat{a}_n^\dagger$ creates one microwave photon with frequency $\omega_n$ in the system. These satisfy the commutation relations $[\hat{a}_j, \hat{a}_k] = [\hat{a}_j^{\dagger}, \hat{a}_k^{\dagger}]=0$ and $[\hat{a}_j, \hat{a}_k^{\dagger}] = \delta_{jk}$.

With these relations, we arrive at the standard quantum Hamiltonian of a multi-mode system,
\begin{align}
\hat{H} = \sum_{n=0}^{\infty} \hbar\omega_n \left(\hat{a}_n^\dagger \hat{a}_n + \frac{1}{2}\right).
\end{align}
The time-dependence of the operators can be obtained from the Heisenberg equation of motion.
Substituting the approximate form of the normal-mode frequency, Eq.~\eqref{eq:mode-freq-optomechanics}, the Hamiltonian becomes
\begin{align}
\hat{H} &= \sum_{n=0}^{\infty} \hbar\omega_n^{(0)}\left[1-(-1)^n \frac{\omega_\mathrm{c}\left(\Delta d^{(1)}\right)^2}{ v_0 D}  {\delta\Phi_\mathrm{ext}^2}\right]\hat{a}_n^\dagger \hat{a}_n, \label{eq:hamiltonian-classical}
\end{align}
where constant terms have been dropped for simplicity. This is the classical quadratic optomechanical Hamiltonian, where the frequency of each eigenmode is a quadratic function of the pseudo-mechanical degree of freedom (flux variation, $\delta\Phi_\mathrm{ext}$).

\subsection{Quantum quadratic optomechanics}\label{sec:quantum-opto}
We denote the capacitively-coupled tunable resonators of Sec.~\ref{sec:analogcircuit} as ``resonator A'', and rewrite the Hamiltonian of Eq.~\eqref{eq:hamiltonian-classical} as $\hat{H}_\mathrm{A}$. We now introduce another uniform TL resonator, denoted as ``resonator B'' (see Fig.~\ref{fig:intro-schematic}). In general, the flux $\hat{\Phi}_\mathrm{B} (z)$ and the Hamiltonian $\hat{H}_\mathrm{B}$ can be written as ($z$ is the new coordinate system describing the resonator B):
\begin{align}
\begin{split}
\hat{\Phi}_\mathrm{B} (z) &= \sum_{m} \sqrt{\frac{\hbar}{2\Omega_m C_{\Sigma,\mathrm{B}}}}u^\mathrm{B}_m(z) \left(\hat{b}_m^\dagger + \hat{b}_m \right),\\
\hat{H}_\mathrm{B} &= \sum_{m} \hbar \Omega_m \hat{b}_m^\dagger \hat{b}_m,
\end{split}\label{eq:resonator-B-fields}
\end{align}
where $C_{\Sigma,\mathrm{B}}$ is the total capacitance, $\Omega_m$ is the mode frequency, and $u^\mathrm{B}_m(z)$ is the mode function of the resonator B, which can be obtained following the procedures used in Sec.~\ref{sec:circuitmodel}. Here, $\hat{b}_m$ and $\hat{b}_m^\dagger$ are the annihilation and the creation operators satisfying the commutation relations $[\hat{b}_j, \hat{b}_k] = [\hat{b}_j^{\dagger}, \hat{b}_k^{\dagger}]=0$ and  $[\hat{b}_j, \hat{b}_k^{\dagger}] = \delta_{jk}$. The annihilation operator $\hat{b}_m$ destroys one microwave photon from the resonator B, whose frequency is $\Omega_m$; the creation operator $\hat{b}_m^\dagger$, on the other hand, creates one microwave photon of frequency $\Omega_m$ in the resonator B.

Now, we assume that the variation in the external flux $\delta\Phi_\mathrm{ext}$ arises from the magnetic field that is generated by the resonator B. In this case, the variation in the external flux becomes a quantum variable $\delta\hat{\Phi}_\mathrm{ext}$, written as \cite{johansson:2014},
\begin{align}
\delta\hat{\Phi}_\mathrm{ext} = \sum_{m} G_m \left(\hat{b}_m^\dagger  + \hat{b}_m \right),\label{eq:quantum-external-flux}
\end{align}
where the coefficients $G_m$ are determined by the experimental configuration. This form can be understood from the fact that the magnetic field is proportional to the current along the resonator B, so that $\delta\hat{\Phi}_\mathrm{ext} \propto \hat{I}_\mathrm{B}(-z_0) \propto \partial_{z} \hat{\Phi}_\mathrm{B} (-z_0)$.

We also assume that the pseudo-mechanical mode frequencies $\Omega_m$ are small compared to the optical mode frequencies $\omega_n$, so that the resonator A adiabatically follows the dynamics of the resonator B.  In this case, the dependence of normal-mode frequency $\omega_n$ on the external flux variation $\delta\Phi_\mathrm{ext}$ is well-defined also for a quantum variable, and we can substitute Eq.~\eqref{eq:quantum-external-flux} into Eq.~\eqref{eq:hamiltonian-classical} with $\delta\Phi_\mathrm{ext} \rightarrow \delta\hat{\Phi}_\mathrm{ext}$. The Hamiltonian of the system consisting of the resonator A and the resonator B then becomes:
\begin{align}
\begin{split}
\hat{H}  	 &= \sum_{n=0}^{\infty}\hbar\omega_n^{(0)} \hat{a}_n^\dagger \hat{a}_n+\sum_{m} \hbar \Omega_m \hat{b}_m^\dagger \hat{b}_m \\
	 &\quad- \sum_{n}^{\infty} \sum_{m, l} \hbar \gamma_{nml}\ \hat{a}_n^\dagger \hat{a}_n \left(\hat{b}_m^\dagger +\hat{b}_m \right)\left(\hat{b}_l^\dagger+\hat{b}_l \right),
\end{split} \label{eq:quad-opto-hamiltonian-multi}
\end{align}
where the coupling tensor $\gamma_{nml}$ is given by
\begin{align}
\gamma_{nml} = (-1)^n \frac{\omega_n^{(0)} \omega_\mathrm{c} \left(\Delta d^{(1)}\right)^2}{v_0 D}G_m G_l.
\end{align}
The tensor $\gamma_{nml}$ quantifies the interaction between three resonator modes: the $n$-th mode of the resonator A, the $m$-th and the $l$-th mode of the resonator B.

The Hamiltonian of Eq.~\eqref{eq:quad-opto-hamiltonian-multi} reduces to the quadratic optomechanical Hamiltonian if we restrict the dynamics to only involve a single mode of each resonator (i.e., by selectively exciting a single mode of each resonator). For instance, by only considering the $n$-th mode of the resonator A and the $m$-th mode of the resonator B, the Hamiltonian takes the standard quadratic optomechanical form:
\begin{align}
\hat{H} = \hbar \omega_n^{(0)} \hat{a}_n^\dagger \hat{a}_n + \hbar \Omega_m \hat{b}_m^\dagger \hat{b}_m - \hbar g_{nm} \hat{a}_n^\dagger \hat{a}_n (\hat{b}_m^\dagger + \hat{b}_m)^2.\label{eq:hamiltonian-quad-opto}
\end{align}
where $g_{nm}\equiv \gamma_{nmm}$ is the quadratic coupling strength of the $n$-th mode of the resonator A and the $m$-th mode of the resonator B. This corresponds to an optical cavity of unperturbed resonance frequency $\omega_n^{(0)}$ coupled to a semi-transparent membrane in the middle, oscillating with mechanical oscillation frequency $\Omega_m$.
The coupling strength $g_{nm}$ can be written as
\begin{align}
g_{nm} =(-1)^n \omega_n^{(0)} G_m^2 \frac{\omega_\mathrm{c}  \left(\Delta d^{(1)}\right)^2}{v_0 D},\label{eq:coupling-strength}
\end{align}
and it follows that $g_{nm} \propto \frac{1}{C_\mathrm{c}}\tan{^2\left(\pi\Phi_\mathrm{ext}^0/{\Phi_0}\right)}$.

Equation~\eqref{eq:coupling-strength} implies that the coupling strength is tunable: in addition to the geometrical arrangement of the system which determines $G_m$, the optomechanical coupling strength can be adjusted by controlling either the capacitive coupling $C_\mathrm{c}$ or the bias flux $\Phi_\mathrm{ext}^0$. The optomechanical coupling is strong when the capacitive coupling is weak ($C_\mathrm{c}\rightarrow 0$) or the bias flux $\Phi_\mathrm{ext}^0$ is close to half-integer multiples of $\Phi_0$ (but not too close to break the $\eta_n^\alpha\rightarrow 0$ assumption). On the contrary, if the capacitive coupling is stronger or the bias flux is closer to integer multiples of a flux quantum, the optomechanical coupling strength decreases.

\section{Circuit realization}\label{sec:circuit-realization}
In this section, we propose a circuit design to realize the quadratic optomechanical Hamiltonian of Eq.~\eqref{eq:hamiltonian-quad-opto}. A detailed analysis on the schematic illustration in Fig.~\ref{fig:intro-schematic} will be presented in Sec.~\ref{sec:circuit-layout}. We discuss which modes of the resonators are suitable for describing the quadratic optomechanical Hamiltonian. In Sec.~\ref{sec:inductive-coupling}, we derive the analytic expression for the coupling constants $G_{m}$ corresponding to inductive coupling between the resonators A and B as in Ref.~\cite{johansson:2014}, and the quadratic optomechanical coupling strength $g_{nm}$ follows. In Sec.~\ref{sec:field-strength}, we suggest a criterion for the field strength of the resonator B in order to retain the quadratic coupling. Estimates on the coupling strength $g_{nm}$ and the upper limit on the field strength will be provided using realistic parameters.

\subsection{Circuit layout}\label{sec:circuit-layout}
\begin{figure}
	\includegraphics[width=0.48\textwidth]{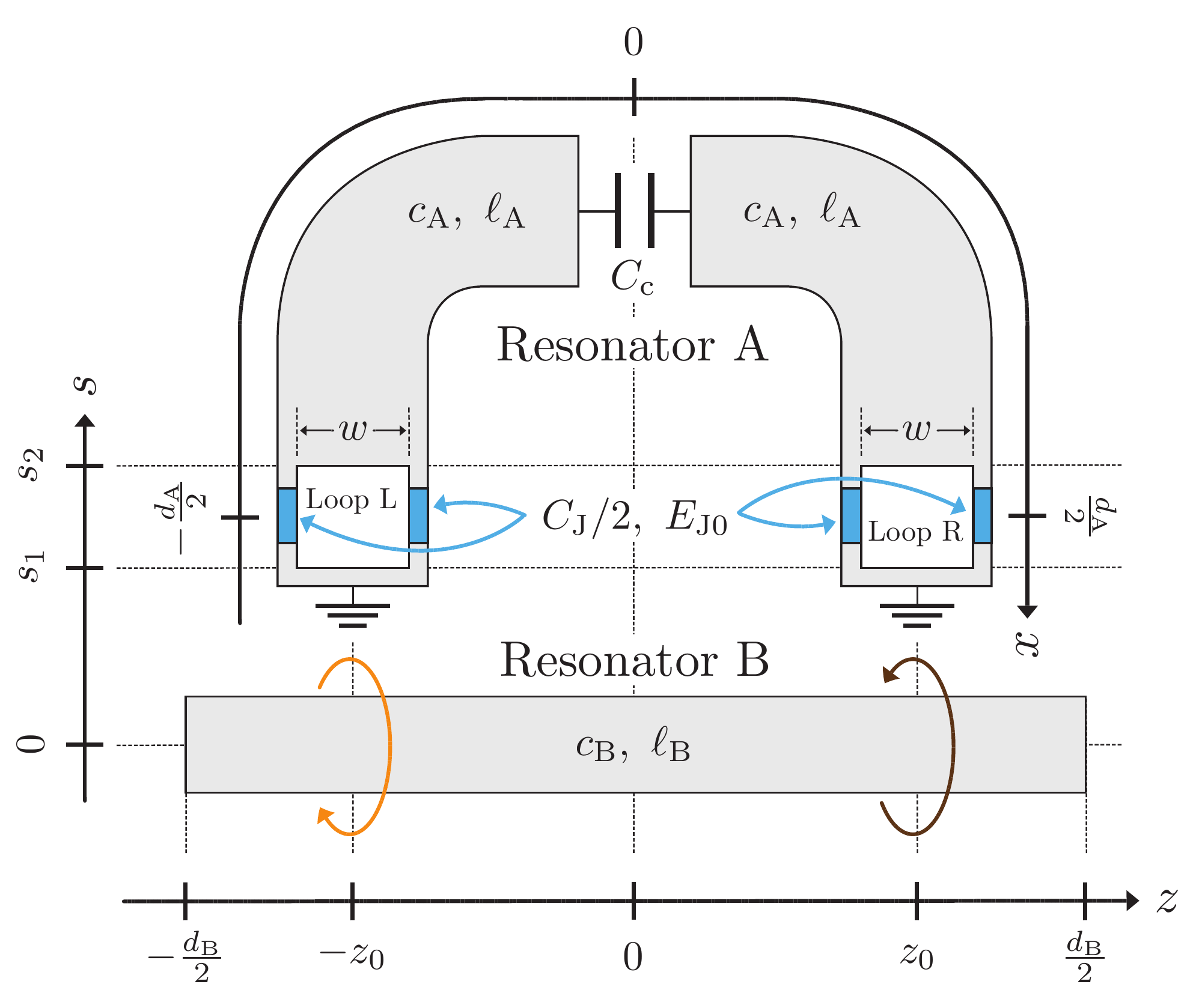}
	\caption{(color online) Detailed layout of Fig.~\ref{fig:intro-schematic}, composed of resonator A (two capacitively-coupled tunable resonators) and resonator B (a TL resonator). The SQUID loops that belong to tunable resonators are denoted as loop L and loop R. The total length, the characteristic capacitance, and inductance per unit length of the resonator $\alpha=\mathrm{A},\ \mathrm{B}$ are given by $d_\alpha$, $c_{\alpha}$, and $\ell_{\alpha}$, respectively. All Josephson junctions are equal with junction capacitance $C_\mathrm{J}/2$ and Josephson energy $E_{J0}$. The three coordinate axes ($x$, $z$, and $s$) specify positions in the system.}
	\label{fig:circuit-realization}
\end{figure}

We investigate the configuration illustrated in Fig.~\ref{fig:circuit-realization}. As in Sec.~\ref{sec:quantum-opto}, two resonators---resonator A and resonator B, which correspond to capacitively-coupled tunable resonators and a TL resonator---are taken into account. Note that the resonator A is bent in such a way that provides an inductive coupling with the resonator B at two sites (loop L and loop R). We assume that the TL resonators forming the resonator $\alpha$ are uniform and have the characteristic capacitance and inductance per unit length $c_{\alpha}$ and $\ell_{\alpha}$. Also, we define $d_\alpha$ as the total length of the resonator $\alpha$ ($\alpha=\mathrm{A},\ \mathrm{B}$). All Josephson junctions in the resonator A are set to have equal junction capacitance $C_\mathrm{J}/2$ and Josephson energy $E_\mathrm{J0}$.

We introduce three coordinate axes ($x$, $z$, and $s$) for the full characterization of the system. The curvilinear coordinate $x$ is the longitudinal coordinate of the resonator A. The two tunable resonators that make up resonator A, each ranging over $-\frac{d_\mathrm{A}}{2} < x < 0$ and $0 < x < \frac{d_\mathrm{A}}{2}$, interact with each other through the capacitor $C_\mathrm{c}$ at $x=0$. The ground-ended SQUIDs of the tunable resonators are placed at $x=\pm\frac{d_\mathrm{A}}{2}$. The linear axes $z$ and $s$ describe the longitudinal and the transverse coordinates of the resonator B, which extends over $-\frac{d_\mathrm{B}}{2} < z < \frac{d_\mathrm{B}}{2}$. The symmetry axis of the system lies at $x=z=0$. The SQUIDs of the resonator A are placed at $z=\pm z_0$ and $s_1 < s < s_2$, with a width $w$ along the $z$-axis. The loops L and R are subject to the equal bias flux $\Phi_\mathrm{ext}^0$.

Following the conventions of Sec.~\ref{sec:quantum-opto}, we denote the normal-mode frequencies of the resonator A and the resonator B as $\omega_n$ and $\Omega_m$. Also, we denote the annihilation operators corresponding to the $n$-th mode of resonator A and the $m$-th mode of resonator B as $\hat{a}_n$ and $\hat{b}_m$.

This system meets the three requirements mentioned in the previous sections: $(i)$ tunable resonators are employed to simulate a cavity whose effective length can be varied by the fluxes threading the SQUID loops; $(ii)$ the tunable resonators are capacitively coupled to each other to introduce reflection and transmission of waves which is similar to a semi-transparent membrane in the middle; $(iii)$ the effective length of the tunable resonators are coupled to the quantum fields $\hat{b}_m$ of a single resonator.

It remains to make sure that the fluxes through the SQUIDs have a variation of the same magnitude in the opposite direction ($\pm\delta\hat{\Phi}_\mathrm{ext}$), in addition to the equal bias flux $\Phi_\mathrm{ext}^0$.
To do so, suppose that we only excite the $m$-th mode of resonator B. Then, the flux field of the resonator B is given by Eq.~\eqref{eq:resonator-B-fields}:
\begin{align*}
\hat{\Phi}_\mathrm{B} (z) = \sqrt{\frac{\hbar}{2\Omega_m c_\mathrm{B} d_\mathrm{B}}} u_m^\mathrm{B} (z) \left(\hat{b}_m^\dagger + \hat{b}_m\right),
\end{align*}
where $u_m^\mathrm{B} (z) $ is the normal-mode function of resonator B. The current $\hat{I}_\mathrm{B} (z) = -\frac{1}{\ell_\mathrm{B}}\partial_z \hat{\Phi}(z)$ along the resonator B at $z=\pm z_0$ is
\begin{align*}
\hat{I}_\mathrm{B} (\pm z_0) = -\frac{1}{\ell_\mathrm{B}}\sqrt{\frac{\hbar}{2\Omega_m c_\mathrm{B} d_\mathrm{B}}} {u_m^\mathrm{B}}' (\pm z_0) \left(\hat{b}_m^\dagger + \hat{b}_m\right).
\end{align*}
The flux variation threading the SQUID loop is proportional to the current along resonator B at $z=\pm z_0$. That is, the flux variations $\delta\hat{\Phi}_\mathrm{ext}$ through the loop L, and $-\delta\hat{\Phi}_\mathrm{ext}$ through the loop R, are proportional to $\hat{I}_\mathrm{B}(-z_0)$ and $\hat{I}_\mathrm{B}(z_0)$, respectively. This requires that the current along the resonator B have opposite signs at $z=\pm z_0$,
\begin{subequations}
\begin{align}
\hat{I}_\mathrm{B}(-z_0) = -\hat{I}_\mathrm{B}(z_0),
\end{align}
or, equivalently, the derivative of the resonator mode function should have opposite signs at $z=\pm z_0$,
\begin{align}
{u_m^\mathrm{B}}'(-z_0) = -{u_m^\mathrm{B}}'(z_0).
\end{align}
\end{subequations}
This is possible when the mode function ${u_m^\mathrm{B}}(z)$ is an even-parity function of $z$, i.e., ${u_m^\mathrm{B}}(z)={u_m^\mathrm{B}}(-z)$. In addition, the anti-node of the mode function should not be located at $z=\pm z_0$, since the anti-nodes correspond to the nodes of the current, where the flux variation is zero.

Therefore, if we are working in a regime where the effective length interpretation in Sec.~\ref{sec:squidres} is valid, i.e., considering low-enough $n$ modes of the resonator A and bias fluxes $\Phi_\mathrm{ext}^0$ not too close to half-integer multiples of a flux quantum to ensure $\eta_n^{\mathrm{L},\mathrm{R}}\rightarrow0$, it is possible to construct the quadratic optomechanical Hamiltonian
\begin{align*}
\hat{H} = \hbar \omega_n^{(0)} \hat{a}_n^\dagger \hat{a}_n + \hbar \Omega_m \hat{b}_m^\dagger \hat{b}_m - \hbar g_{nm} \hat{a}_n^\dagger \hat{a}_n \left(\hat{b}^\dagger_m +\hat{b}_m\right)^2,
\end{align*}
by considering the resonator B mode functions with even parity. Here, the unperturbed normal-mode frequency $\omega_n^{(0)}$ of the resonator A, and the quadratic coupling strength $g_{nm}$ between the $n$-th mode of the resonator A and $m$-th mode of the resonator B is obtained from Eq.~\eqref{eq:unperturbed-normal-mode-freq} and Eq.~\eqref{eq:coupling-strength}, with redefinition of parameters $\ell_0 \rightarrow \ell_\mathrm{A}$, $c_0 \rightarrow c_\mathrm{A}$, and $d\rightarrow d_\mathrm{A}$,
\begin{align}
\begin{split}
\omega_n^{(0)} &= \frac{\pi v_{\mathrm{A}}}{D_\mathrm{A}}\bigg( n+\textrm{mod}(n+1,2)\bigg) \\
			&\quad -\frac{2v_{\mathrm{A}}\cos{^{-1}(|r_n^{(0)}|)}}{D_\mathrm{A}}\textrm{mod}(n+1,2),
\end{split}\\
g_{nm} &= (-1)^n\frac{ \omega_n^{(0)} c_\mathrm{A}}{C_\mathrm{c} D_\mathrm{A}}\left(\frac{G_m \Phi_0}{4\pi\ell_\mathrm{A} E_{\mathrm{J}0}}\right)^2 \tan{^2\left(\pi\frac{\Phi_\mathrm{ext}^0}{\Phi_0}\right)} ,\label{eq:coupling-strength-1}
\end{align}
where $v_{\mathrm{A}} = 1/\sqrt{\ell_{\mathrm{A}}c_{\mathrm{A}}}$ is the velocity of the wave inside the resonator A, and $D_\mathrm{A} = d_\mathrm{A} + 2 \left(\frac{\Phi_0}{2\pi}\right)^2 ({\ell_{\mathrm{A}}E_{J0}})^{-1}$ is the total effective length of the resonator A in the absence of the flux variation. Here, $|r_n^{(0)}|$ is the absolute value of the reflectivity arising from the capacitive coupling at $x=0$ and can be obtained from Eq.~\eqref{eq:reflectivity-absolute}.

In particular, we consider the case where the resonator B is open-ended, i.e., $\partial_z \hat{\Phi}_\mathrm{B} (\pm\frac{d_\mathrm{B}}{2}) = 0$. Then, the normal-mode frequency $\Omega_m$ and the normal-mode function $u_m^\mathrm{B}(z)$ is given by ($m=1,2,\ldots$)
\begin{align}
\Omega_m &= \frac{m\pi v_\mathrm{B}}{d_\mathrm{B}},\\
u_m^\mathrm{B} (z) &=  \left\{\begin{alignedat}{2}
				    &\sqrt{2}\sin{\left(\frac{m\pi z}{d_\mathrm{B}}\right)}\quad &(m:& \textrm{ odd}),\\
				    &\sqrt{2}\cos{\left(\frac{m\pi z}{d_\mathrm{B}}\right)}&(m:& \textrm{ even}).
			\end{alignedat}\right.
\end{align}
Here, even and odd values of $m$ correspond to even-parity mode functions and odd-parity mode functions, respectively. We conclude that sufficiently low modes of the resonator A, together with even modes ($m=2,\ 4,\ \ldots$) of the resonator B, are plausible candidates for the circuit realization of quadratic optomechanics.

\subsection{Inductive coupling}\label{sec:inductive-coupling}

In this section, we discuss the inductive coupling of the resonator A and the resonator B. The coefficient $G_m$ relating the effective displacement parameter $\hat{\xi}$ and the flux variation $\delta\hat{\Phi}_\mathrm{ext}$ can be obtained by considering the geometrical configuration of the system \cite{johansson:2014}.

The magnetic field at $(z,s)$ generated by the current distribution $\hat{I}_\mathrm{B} (z)$ of resonator B is estimated from the Biot-Savart law,
\begin{align}
\hat{B}(z, s) = \frac{\mu_0 }{4\pi} \int_{-d_\mathrm{B}/2}^{d_\mathrm{B}/2} \frac{s \hat{I}_\mathrm{B} (z') \: \mathrm{d} z'}{\left[s^2 + (z-z')^2\right]^{3/2}},\label{eq:biot-savart}
\end{align}
where $\mu_0 = 4\pi\times 10^{-7}\ \mathrm{H\ m^{-1}}$ is the permeability of free space. Note that the magnetic field is described as a quantum operator. If the point in consideration is sufficiently close to the resonator B, compared to its dimension, i.e., $s \ll d_\mathrm{B}$, the integrand of Eq.~\eqref{eq:biot-savart} contributes significantly only in the range $|z'-z|\le s$, and the limits of integration $\pm d_\mathrm{B}/2$ can be replaced with $\pm \infty$. Also, if the variation in the current distribution $\hat{I}_\mathrm{B}(z')$ is negligible near $z'=z$, then Eq.~\eqref{eq:biot-savart} can be approximated as
\begin{align*}
\hat{B}(z, s)  \approx \frac{\mu_0 s}{4\pi} \hat{I}_\mathrm{B} (z) \int_{-\infty}^{\infty} \frac{s \:\mathrm{d} z'}{\left[s^2 + (z-z')^2\right]^{3/2}} 
	= \frac{\mu_0 \hat{I}_\mathrm{B} (z)}{2\pi s},
\end{align*}
which is the magnetic field arising from a straight wire carrying a constant current. Thus, if the SQUIDs are placed very close to resonator B ($s_1, s_2\ll d_\mathrm{B}$), and the positions of the SQUIDs $z=\pm z_0$ correspond to nodes of the normal-mode function $u_m^\mathrm{B}(z)$ (the variation in the current distribution is minimal), the flux variation $\delta\hat{\Phi}_\mathrm{ext}$ through the SQUIDs are
\begin{align*}
\delta\hat{\Phi}_\mathrm{ext} \approx w\int_{s_1}^{s_2} \hat{B}(-z_0, s)\: \mathrm{d} s = G_m\left(\hat{b}^\dagger_m +\hat{b}_m\right),
\end{align*}
where the inductive coupling coefficient $G_m$ is given by 
\begin{align}
G_m = \pm \frac{\mu_0 w}{2\pi \ell_\mathrm{B} d_\mathrm{B}} \sqrt{\frac{m\pi\hbar}{v_\mathrm{B} c_\mathrm{B}}} \ln{\left(\frac{s_2}{s_1}\right)} .\label{eq:inductive-coupling-coeff}
\end{align}
Here, the sign of the coefficient depends on which node of the normal-mode function we choose as $z=z_0$. If we further assume that the dimension of the SQUID is much smaller than its distance from resonator B, i.e., $(s_2 - s_1) \ll s_1$, we can apply the approximation
\begin{align*}
\ln{(1+x)}\approx x\quad(|x|\ll1),
\end{align*}
to simplify Eq.~\eqref{eq:inductive-coupling-coeff}:
\begin{align}
G_m = \pm \frac{\mu_0}{2\pi \ell_\mathrm{B}} \frac{A}{d_B s_1}\sqrt{\frac{m\pi\hbar}{v_\mathrm{B} c_\mathrm{B}}}.\label{eq:inductive-coupling-coeff-simp}
\end{align}
Here, $A\equiv w(s_2-s_1)$ is defined as the area enclosed by the SQUID loop. Combining Eqs.~\eqref{eq:coupling-strength-1} and \eqref{eq:inductive-coupling-coeff-simp}, the ratio of the coupling strength $g_{nm}$ to the product of normal-mode frequencies $\omega_n^{(0)}\Omega_m$, is written as ($n=0, 1, 2, \ldots$ and $m = 2, 4, 6, \ldots$)
\begin{align}
\begin{split}
\frac{\hbar g_{nm}}{(\hbar\omega_n^{(0)})(\hbar\Omega_m)} &= (-1)^n \frac{\ell_\mathrm{B} d_\mathrm{B}}{\Phi_0^2} \left(\frac{c_\mathrm{A} D_\mathrm{A}}{C_\mathrm{c}}\right) \left(\frac{L_{\mathrm{J}0}}{\ell_\mathrm{A} D_\mathrm{A}}\right)^2 \\ &\quad \times\left(\frac{A}{d_\mathrm{B} s_1}\right)^2 \left(\frac{\mu_0}{\ell_\mathrm{B}}\right)^2  \tan{^2 \left(\pi\frac{\Phi_\mathrm{ext}^0}{\Phi_0}\right)},\end{split}
\label{eq:relative-coupling-strength}
\end{align}
where $L_{\mathrm{J}0}$ is the effective inductance of the SQUID in the absence of the external flux, $L_{\mathrm{J}0} = \frac{1}{2E_{\mathrm{J}0}}\left(\frac{\Phi_0}{2\pi}\right)^2$.  From the discussions of Sec.~\ref{sec:squidres}, it is important that $L_{\mathrm{J}0}/\ell_{\mathrm{A}}D_{\mathrm{A}}$ be smaller than $10^{-2}$ and that $\Phi_\mathrm{ext}^0$ not be too close to half-integral multiples of $\Phi_0$ in order to maintain the effective length interpretation. Note that the absolute value of this ratio is independent of $n$ and $m$.

\begin{figure}
  \includegraphics[width=0.48\textwidth]{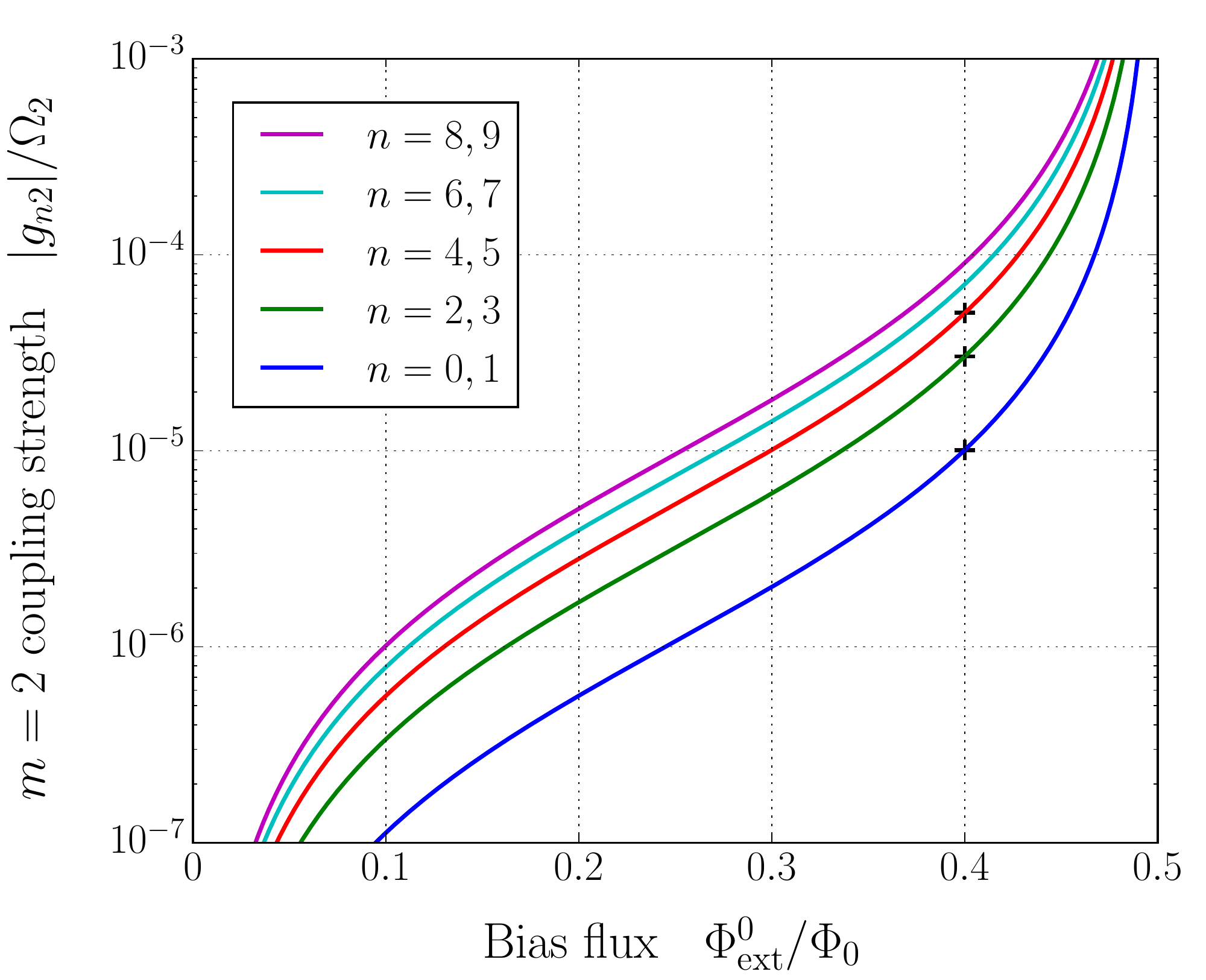}
  \caption{(color online) The normalized coupling strength $|g_{nm}|/\Omega_{m}$ as a function of the bias flux, for $m=2$ and different $n$ modes ($n=0$ to $n=9$, from bottom to top). The parameters $d_\mathrm{A} = d_\mathrm{B}/20 = 20\ \mathrm{mm}$, $A/d_\mathrm{B}s_1 = 10^{-3}$, $\ell_{\mathrm{A}}=\ell_\mathrm{B}=4.57\times 10^{-7}\ \mathrm{H\ m^{-1}}$, $c_\mathrm{A}=c_\mathrm{B} = 1.46\times 10^{-10}\ \mathrm{F\ m^{-1}}$, $C_c = 1\ \mathrm{fF}$, $C_\mathrm{J} = 30\ \mathrm{fF}$, and $E_{\mathrm{J}0} = 6.17\times 10^{-22}\ \mathrm{J}$ were used to evaluate Eq.~\eqref{eq:relative-coupling-strength}. The two modes labeling a single line ($n=8,9$ and the uppermost line, for instance) are in fact different but too close to be distinguishable in the plot. The three markers, \textbf{\textsf{+}}, at $\Phi^0_\mathrm{ext}/\Phi_0=0.4$ correspond to those in Fig.~\ref{fig:coupling-strength-capacitance}.\\}
	\label{fig:coupling-strength-estimates}

  \includegraphics[width=0.48\textwidth]{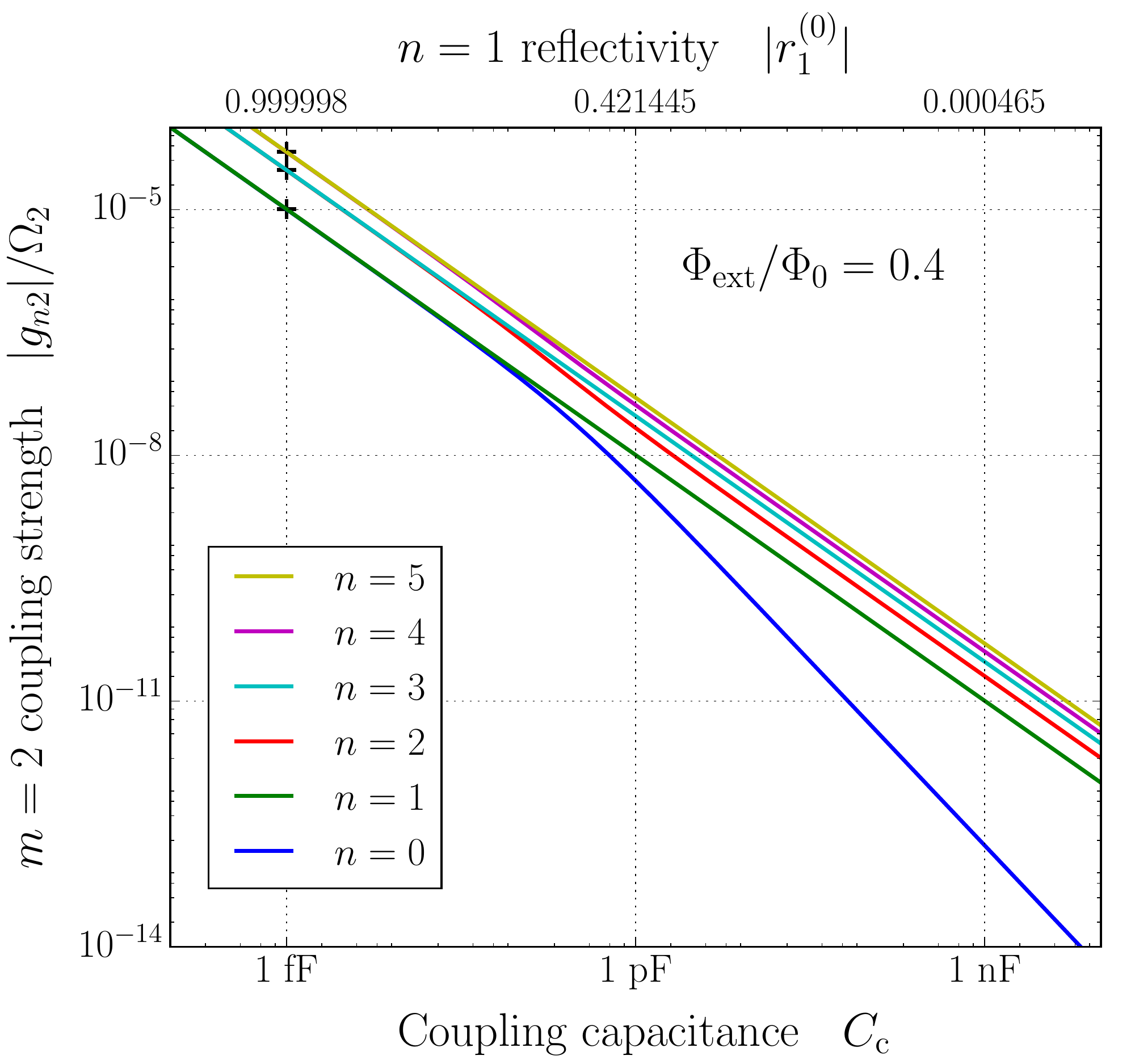}
  \caption{(color online) The normalized coupling strength  $|g_{nm}|/\Omega_{m}$ as a function of the coupling capacitance $C_\mathrm{c}$ of resonator A, for $m=2$ and different $n$ modes ($n=0$ to $n=5$, from bottom to top). All parameters used are the same as in Fig.~\ref{fig:coupling-strength-estimates}, except for the fixed bias flux $\Phi^0_\mathrm{ext}/\Phi_0 = 0.4$ and varying capacitance $C_\mathrm{c}$. The three markers, \textbf{\textsf{+}}, at $C_\mathrm{c}=1\ \mathrm{fF}$ correspond to those in Fig.~\ref{fig:coupling-strength-estimates}. The upper horizontal axis is the reflectivity of $n=1$ mode obtained from $C_\mathrm{c}$.}
  \label{fig:coupling-strength-capacitance}
\end{figure}

The ratio is inversely proportional to the coupling capacitance $C_\mathrm{c}$ and depends on the bias flux with $\tan{^2\left(\pi\Phi_\mathrm{ext}^0/\Phi_0\right)}$ as discussed in Eq.~\eqref{eq:coupling-strength}. Also, a geometric factor $A/d_\mathrm{B} s_1$ is involved in the expression, with a quadratic dependence. This is due to the fact that the flux through the SQUID loop, which is dependent on the area enclosed by the loop and the distance from the current source, plays a significant role in the pseudo-mechanical coupling.

We define the normalized coupling strength as the ratio of the coupling strength $g_{nm}$ to the mode frequency $\Omega_{m}$ of the resonator B. In Figs.~\ref{fig:coupling-strength-estimates}-\ref{fig:coupling-strength-capacitance}, the normalized coupling strengths are illustrated as a function of the bias flux $\Phi_\mathrm{ext}^0/\Phi_0$ and coupling capacitance $C_\mathrm{c}$. Realistic parameters, which yields $L_{\mathrm{J}0} < 10^{-2}\ \ell_{\mathrm{A}}D_\mathrm{A}$ and $C_\mathrm{J} < 10^{-2}\ c_\mathrm{A} D_\mathrm{A}$, have been used to evaluate Eq.~\eqref{eq:relative-coupling-strength}.

In Fig.~\ref{fig:coupling-strength-estimates}, the low capacitive coupling regime (equivalently, the high reflectivity regime) is taken into account. The coupling strength is larger for higher resonator A modes and smaller for lower resonator A modes. The coupling strength grows infinitely high as the bias flux $\Phi_\mathrm{ext}^0$ approaches $0.5\ \Phi_0$. In particular, for the bias flux of $\Phi_\mathrm{ext}^0 /\Phi_0= 0.4$ and $n=1$, which is in the regime where the effective length interpretation is valid, the normalized coupling strength has the value $g_{12}/\Omega_2 \approx 10^{-5}$. This is approximately \emph{five orders of magnitude higher} than the normalized quadratic coupling strength Eq.~\eqref{eq:coupling-strength-cavity-2} in the cavity optomechanical system of Ref.~\cite{Sankey:2010ej}.

Figure~\ref{fig:coupling-strength-capacitance} describes the dependence of the normalized coupling strength on the coupling capacitance $C_\mathrm{c}$, for a fixed value of bias flux $\Phi_\mathrm{ext}^0 /\Phi_0= 0.4$. The capacitance $C_\mathrm{c}$ and the absolute value of the effective reflectivity $|r_n^{(0)}|$ are converted to each other according to Eq.~\eqref{eq:reflectivity-absolute}. As the capacitance becomes smaller, the normalized coupling strength $g_{nm}/\Omega_m$ increases, and vice versa. Note that each seemingly degenerate mode of $C_\mathrm{c}=1\ \mathrm{fF}$ are resolved into two distinct modes as the coupling capacitance $C_\mathrm{c}$ grows. In the low $C_\mathrm{c}$ regime, the two tunable resonators forming resonator A are almost decoupled, and the deviation from the degeneracy point is very small. However, as the capacitance $C_\mathrm{c}$ grows, this deviation becomes larger, showing significant differences between modes.

\subsection{Field strength}\label{sec:field-strength}
For the system to retain a quadratic coupling, there is a restriction on the expectation value and the fluctuations in the flux variation, $\langle\delta\hat{\Phi}_\mathrm{ext}\rangle$ and $\Delta(\delta\hat{\Phi}_\mathrm{ext})\equiv \sqrt{\langle\delta\hat{\Phi}_\mathrm{ext}^2\rangle-\langle\delta\hat{\Phi}_\mathrm{ext}\rangle^2}$. This is due to the fact that the displacement parameter $ \hat{\xi}=\Delta d^{(1)}\delta\hat{\Phi}_\mathrm{ext} $ should lie within a certain range to maintain the quadratic approximation Eq.~\eqref{eq:mode-freq-optomechanics}. Defining the position quadrature of the $m$-th mode of the resonator B as $\hat{X}_m \equiv \hat{b}_m^\dagger + \hat{b}_m = \delta\hat{\Phi}_\mathrm{ext} / G_m$, the criterion becomes:
\begin{align}
\left\vert \langle \hat{X}_m \rangle  \pm \Delta(\hat{X}_m) \right\vert \le X_{nm*},\label{eq:validity-criterion}
\end{align}
where $X_{nm*}(\Phi_\mathrm{ext}^0)\equiv{\xi_{n*}}/{\vert\Delta d^{(1)} G_m\vert}$ is the maximal amplitude of the quadrature $\hat{X}_m$ to maintain a quadratic coupling. Here, $\xi_{n*}$ is the validity extent of the $n$-th mode of the resonator A, which is obtained from Eq.~\eqref{eq:validity-range}. The fluctuation in the $\hat{X}_m$ can be explicitly written as
\begin{align}
\begin{split}
\Delta(\hat{X}_m) &= \bigg[\left(\langle\hat{b}_m^{\dagger 2}\rangle -  \langle\hat{b}_m^\dagger\rangle^2\right) + 
\left(\langle\hat{b}_m^{2}\rangle -  \langle\hat{b}_m\rangle^2\right)\\
&\qquad +2\left(\langle\hat{b}_m^\dagger \hat{b}_m\rangle - \langle \hat{b}_m^\dagger \rangle\langle \hat{b}_m \rangle\right) + 1 \bigg]^{1/2}.
\end{split}
\end{align}
Hereafter, we refer to $X_{nm*}$ as the \emph{maximal amplitude}. Note that the maximal amplitude is dependent on the modes $n$ and $m$ of the resonators A and B as well as the bias flux $\Phi_\mathrm{ext}^0$.

\begin{figure}
	\includegraphics[width=0.48\textwidth]{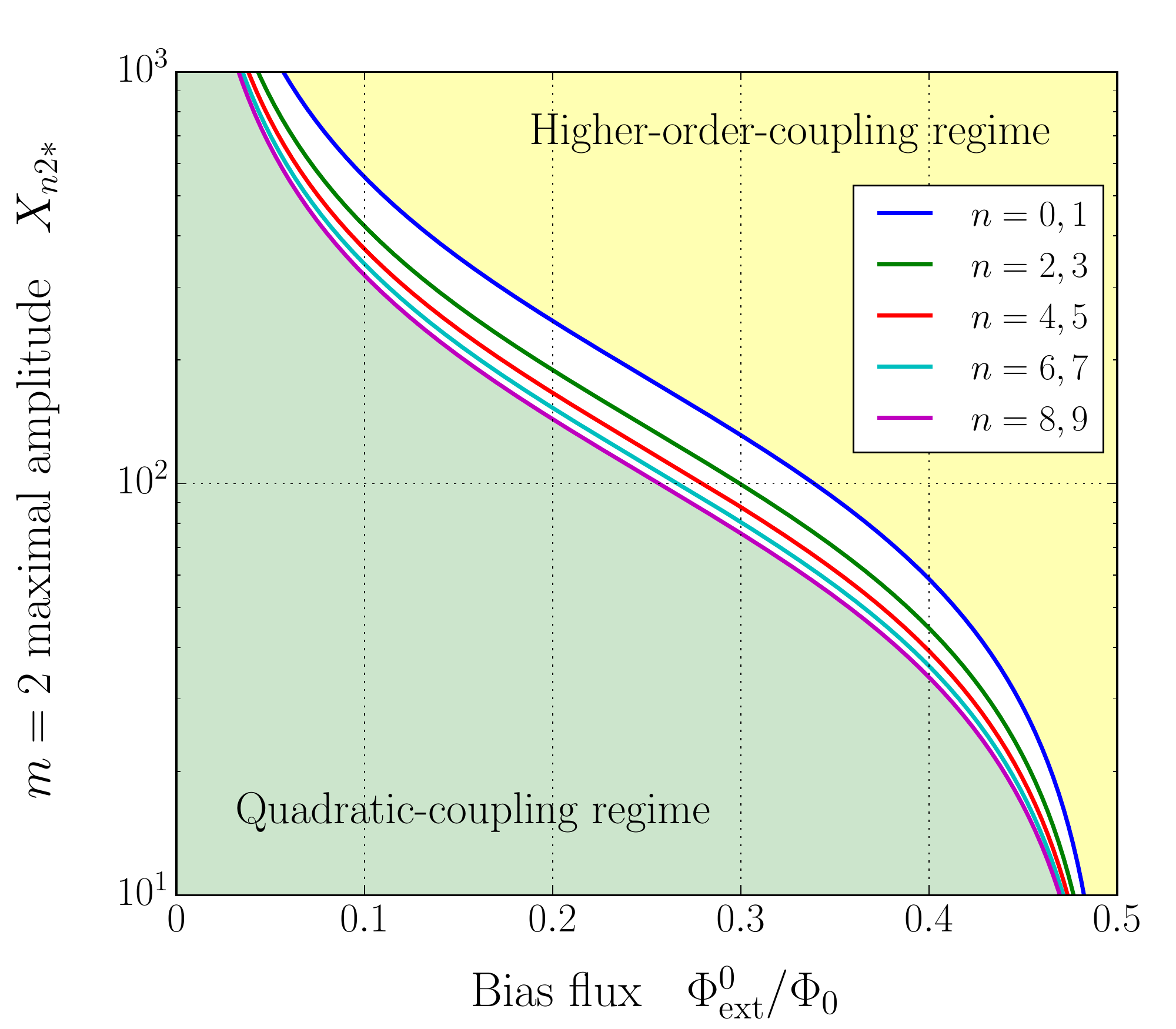}
	\caption{(color online) Maximal amplitudes $X_{nm*}$ of the position quadrature defined in Eq.~\eqref{eq:validity-criterion}, as a function of bias flux $\Phi_\mathrm{ext}^0$, for $m=2$ and different values of $n$. All the parameters used are the same as in Fig.~\ref{fig:validity-extent-estimates}. Each curve corresponds to a discrete $n$ mode. Note that the two modes labeling a single curve ($n=0,1$ and the uppermost curve, for instance) are in fact different but so close to each other as to look degenerate when the reflectivity is high. The upper-right region (yellow) and the lower-left region (green) correspond to the higher-order-coupling regime and the quadratic-coupling regime, respectively.}
	\label{fig:validity-extent-estimates}
\end{figure}

Figure~\ref{fig:validity-extent-estimates} shows the estimates for the maximal amplitude for $m=2$ as a function of bias flux based on the realistic parameters used in Fig.~\ref{fig:coupling-strength-estimates}. Higher $n$ and $m$ modes have lower maximal amplitudes, decreasing by a small amount. The maximal amplitudes are highly affected by the bias flux, especially near half-integral multiples of $\Phi_0$. Using Fig.~\ref{fig:validity-extent-estimates}, we test the validity of the quadratic approximation based on three typical quantum states: the vacuum state, a thermal state, and a coherent state.

\subsubsection{Vacuum state}
For the vacuum state $\left\vert 0 \right\rangle$, the expectation value of the position quadrature is zero, i.e., $\langle\hat{X}_m\rangle = 0$, and only the fluctuations remain. The fluctuations of the position quadrature for the vacuum state are given by $\Delta (\hat{X}_m) = 1$. Thus, the criterion of Eq.~\eqref{eq:validity-criterion} reduces to the inequality $X_{nm*}\ge 1$. For the settings in Fig.~\ref{fig:validity-extent-estimates}, this inequality is readily satisfied unless the bias flux approaches half-integral multiples of a flux quantum within $\Phi_0/250$. Thus, the vacuum fluctuations lie well inside the quadratic coupling regime.

\subsubsection{Thermal state}
For a thermal state at temperature $T$, the expectation value and fluctuations of the position quadrature are
\begin{align*}
 \langle\hat{X}_m\rangle =0,\quad \Delta (\hat{X}_m) = \sqrt{\coth{\left(\frac{\hbar\Omega_m}{2k_\mathrm{B}T}\right)}},
 \end{align*}
and the criterion Eq.~\eqref{eq:validity-criterion} reduces to the inequality,
\begin{align*}
\coth{\left(\frac{\hbar\Omega_m}{2k_\mathrm{B}T}\right)}\le (X_{nm*})^2.
\end{align*}
From this, we can obtain upper bounds on the average photon number $\bar{n}= \langle\hat{b}_m^\dagger \hat{b}_m\rangle$ of a thermal state. The condition is given by:
\begin{align*}
\bar{n}= \left[{\exp{\left(\frac{\hbar\Omega_m}{k_\mathrm{B} T}\right)}-1}\right]^{-1} \le \frac{(X_{nm*})^2 - 1}{2}.
\end{align*}
For a bias flux of $\Phi_\mathrm{ext}^0/ \Phi_0= 0.4$ in Fig.~\ref{fig:validity-extent-estimates}, $X_{nm*}$ for $n=9$ and $m=2$ is approximately $33.8$ and it follows that the upper bound on the average photon number is $\bar{n}\lesssim 572$. 

\subsubsection{Coherent state}
For a time-evolving coherent state $\left|\beta, t\right\rangle =\left| \beta e^{-i\Omega_m t}\right\rangle$, neither the expectation value nor the fluctuations of the position quadrature vanish, and are given by
\begin{align*}
\langle\hat{X}_m\rangle = 2|\beta| \cos{(\Omega_m t-\varphi)},\quad \Delta(\hat{X}_m) = 1,
\end{align*}
where $\varphi$ is the phase defined as $\beta = |\beta| e^{i\varphi}$. Then, the criterion Eq.~\eqref{eq:validity-criterion} reduces to the inequality,
\begin{align*}
2|\beta| + 1 \le X_{nm*}.
\end{align*}
The upper bound on the average photon number is expressed as,
\begin{align*}
\bar{n} = |\beta|^2 \lesssim \frac{(X_{nm*}-1)^2}{4}.
\end{align*}
Thus, for a bias flux of $\Phi_\mathrm{ext}^0/ \Phi_0= 0.4$ in Fig.~\ref{fig:validity-extent-estimates}, it follows that $\bar{n}\lesssim 270$.

\section{Conclusions}\label{sec:conclusion}
In conclusion, we have introduced and analyzed a cQED setup for simulating membrane-in-the-middle optomechanical systems. Two capacitively-coupled SQUID-terminated TL resonators (resonator A) inductively coupled to a TL resonator (resonator B) were used to generate a quadratic-optomechanical-like coupling. A complete description of the Hamiltonian formulation as well as the canonical quantization procedure are provided. Although not discussed explicitly, by introducing an asymmetry in our circuit, either by applying unequal bias fluxes through the SQUIDs or moving the position of the coupling capacitor of resonator A, our circuit enters the standard linear optomechanics regime. Using realistic parameters, the ratio of the quadratic coupling strength to the pseudo-mechanical oscillation frequency is estimated as $10^{-5}$. We note that our proposal anticipates a significant improvement in the quadratic coupling strength to \emph{five orders of magnitude}, from the cavity-optomechanical systems of Refs.~\cite{Thompson:2008dx, Jayich:2008iz, Sankey:2009vs, Sankey:2010ej}.

In general, the superconducting TL resonators could be manufactured with quality factors of $10^{4}$ or higher \cite{Sandberg:2008br, Goppl:2008iu}, and the quadratic coupling strength compared to dissipation rates $\kappa_\mathrm{A}$, $\kappa_\mathrm{B}$ of resonators could be raised to $g/\kappa_\mathrm{A},\ g/\kappa_\mathrm{B} > 0.1$ in our setup. This suggests that the strong-coupling regime of quadratic optomechanics might be achievable, and that our setup would be a good testing ground for quantum phenomena in this regime, e.g., QND measurements \cite{Braginsky:1980qv, Miao:2009jk} of pseudo-mechanical phonon number.

\section*{Acknowledgements}
This work was partly supported by the RIKEN iTHES Project, MURI Center for Dynamic Magneto-Optics, JSPS-RFBR No.~12-02-92100, and a Grant-in-Aid for Scientific Research (S). E.-J. Kim was partly supported by the undergraduate research intership program of College of Natural Sciences, Seoul National University.
\bibliography{references}

\end{document}